\documentclass[11pt,a4paper]{article}

\usepackage[utf8]{inputenc}
\usepackage[T1]{fontenc}
\usepackage[english]{babel}

\usepackage{amsmath,amssymb,amsthm}
\usepackage{stmaryrd}

\usepackage{graphicx}
\usepackage[dvipsnames]{xcolor}
\usepackage{tikz}
\usetikzlibrary{arrows.meta,patterns,calc}

\usepackage{a4wide}
\usepackage{hyperref}

\usepackage{authblk}

\newcommand{\R}{\mathbb{R}}
\newcommand{\ubar}[1]{\underline{#1}}
\newcommand{\llb}{\llbracket}
\newcommand{\rrb}{\rrbracket}

\DeclareMathOperator{\rank}{rank}

\DeclareMathOperator{\dist}{dist}
\DeclareMathOperator{\Ima}{Im}
\DeclareMathOperator{\spann}{span}
\DeclareMathOperator{\idty}{Id}

\newtheorem{theorem}{Theorem}
\newtheorem{lemma}[theorem]{Lemma}
\newtheorem{proposition}[theorem]{Proposition}

\theoremstyle{definition}
\newtheorem{definition}[theorem]{Definition}

\theoremstyle{remark}
\newtheorem{remark}[theorem]{Remark}

\title{Controlled Excitation for Identification of Nonlinear Systems
with Saturated Outputs
\thanks{This work was partially supported by ANR-20-CE48-0003 and ANR-25-CE48-2184.}}
\author[1]{Adel M. Annabi}
\author[1]{Jean-Baptiste Pomet}
\author[2]{Dario Prandi}
\author[1]{Ludovic Sacchelli}

\affil[1]{Université Côte d'Azur, Inria, CNRS, LJAD, France\\
\texttt{\{adel-malik.annabi,jean-baptiste.pomet,ludovic.sacchelli\}@inria.fr}}

\affil[2]{Université Paris-Saclay, CNRS, CentraleSupélec,
Laboratoire des Signaux et Systèmes, 91190 Gif-sur-Yvette, France\\
\texttt{dario.prandi@centralesupelec.fr}}

\date{}

\begin{document}

\maketitle

\begin{abstract}
    We consider nonlinear systems whose measured output is the product of an unknown matrix and a known saturation function of the unmeasured state.
    Generating excitation that allows the matrix to be recovered from the output is challenging when the dynamics are largely unknown.
    We exploit saturation to construct this excitation.
    Under a geometric accessibility condition, we derive explicit open-loop controls that drive the state into regions where the saturation takes prescribed values.
    For componentwise saturations, this condition holds generically with as few as two independent control directions.
    The controls require no state measurement and only partial knowledge of the dynamics, with no information on the structure of its nonlinear part.
    The resulting regressor is known in advance, and the unknown matrix is recovered exactly by a closed-form regression.
    Under stronger structural assumptions and with sufficiently many independent outputs, an output feedback tracks prescribed reference trajectories within the unsaturated region to arbitrary accuracy and permits exact state recovery.
    Suitably chosen references then also enable exact identification of linearly parameterized (nonlinear) dynamics.
    We analyze robustness to measurement noise and illustrate the method numerically in the presence of both measurement and dynamical noise.
\end{abstract}

\section{Introduction}

Saturations and other sigmoidal activation functions occur in many controlled systems.
In neuroscience they are part of the standard models of population activity, such as the Wilson--Cowan equations \cite{Wilson1972,wilson1973mathematical,Destexhe2009}, Amari neural fields \cite{amari1977dynamics} and neural mass models \cite{Jansen1995,david2003neural}.
In these models the saturation converts the mean membrane potential of a population into a firing rate \cite{coombes2014neural,bressloff2011spatiotemporal}.
Although typical saturations are smooth, piecewise linear saturations, reaching their bounds at finite levels, are often used in practice since they make the models more tractable \cite{coombes2018networks}.
We refer to \cite{breakspear2017dynamic} for a review of such models at the scale of the whole brain.

Control theory has been applied to these models for brain monitoring \cite{sood2016nirsParameterEst}, brain--machine interfaces \cite{sorrell2021brainBMI} and closed-loop deep brain stimulation in Parkinson's disease \cite{carron2013closing,detorakis2015closed}.
All of these applications need a model whose parameters are known.
Fitting the parameters offline runs into degeneracy, since distinct parameter sets can reproduce the same data equally well \cite{bernard2023brain}, and it cannot follow parameters that drift over time \cite{sorrell2021brainBMI}.
In such models, moreover, the recorded signals are often weighted sums of the activity of several populations, which depend on the recording device and on the anatomy: this paper is about the identification of such weights.

More precisely, in this work we consider input--output systems of the form
\begin{equation} \label{eq: intro model}
    \dot x = f(x,u,t), \qquad y = W S(x),
\end{equation}
where the dynamics of the state $x\in\R^n$ is governed by the control $u\in\R^m$ via the map $f$, which is largely unknown;
the output $y\in\R^q$ is measured (but not the full state $x$);
$S$ is a known map $\R^n\to\R^n$ assumed constant on some regions (we call such a map a saturation function, see Definition~\ref{def: saturation});
and $W$ is an unknown constant $q\times n$ matrix, whose determination from the measurement of $y$ on a finite time interval is one of the main goals of the paper.
We stress that the control $u$ is left free, and a particularity of our construction is to design a control to ease identification.

This problem is an example of parameter identification 
\cite{ljung1999system,SoderstromStoica1989}. 
For controlled systems, it can usually be solved by a combination of two ingredients: a persistent excitation condition \cite{bazanella2010closed}, and an observer or a contraction property of the dynamics \cite{elliott1979parameter,besanccon2007nonlinear} in order to fit the input-output mapping to the observed data. See, e.g., \cite{brivadis2024adaptive} for a refined example of such principles in the context of dynamics with saturation. These tools, however, do not apply to \eqref{eq: intro model}.

In this
formulation, the unknown matrix $W$ appears multiplied by the time-varying signal $S(x(\cdot))$,
that cannot in general play the role of a regressor\footnote{By regressor we intend, following \cite[Sec.\ 7.3]{ljung1999system}, the quantity $\varphi$ appearing in an equality of the type $y(t)=\varphi(t)\theta$, on which many common identification methods are based.};
it is indeed not measured and cannot be computed, since neither $x$ nor $f$ is fully known, and thus no excitation condition on it can be checked a priori, or even formulated in terms of the data.
Observers do not help, as the contraction, passivity or observability properties they need are not available for $f$ in general.
Relaxations of persistent excitation, such as excitation over a finite interval \cite{kreisselmeier1990richness}, initial excitation \cite{roy2018combined}, concurrent learning \cite{chowdhary2010concurrent} and dynamic regressor extension and mixing \cite{aranovskiy2017parameters},  are not available.
They weaken the requirement on the regressor, but still need it to be known or computable along the trajectory.
The same holds for input design methods for nonlinear identification \cite{Vincent-Novara-Hsu-al2010,Novara-Vincent-Hsu-al2011}, which assume that the regressor, or at least its structure, is available.

This is the typical situation of the high-uncertainty models of neuroscience, in which only the structure of the equations is known.
State and parameter estimation for neural models has been addressed with Kalman filtering \cite{schiff2008kalman,freestone2011data}, with observers for neural mass models \cite{chong2012robust,postoyan2012parameter,lopez2015state}, and with adaptive observers for networks and neural fields \cite{brivadis2022online,burghi2022distributed,burghi2023adaptive,brivadis2024adaptive,annabi2024homogeneous,avitabile2025state}.
These methods require either a detailed model of the dynamics or contraction and persistent excitation assumptions.
Such assumptions are hard to check, and they are only weakly satisfied when saturation decouples some of the variables \cite{annabi2025identification}.
Observe that this question also arises for artificial neural networks, whose weight matrices one may wish to reconstruct from input--output data \cite{Aoyama-Ichikawa}.

We propose an approach that exploits the specific features of saturation nonlinearities.
Indeed, a saturation is constant on large regions of the state space, so the value of the regressor $S(x(\cdot))$ is known whenever the state is known to lie in one of these regions.
In particular, if the control can steer in open loop the state into these regions (say, where $S(x)=v_i$ for known vectors $v_i$) and keep it there for some time, the output is $y=Wv_i$ on known intervals.
Then, $W$ follows from a linear regression whose regressor is known before the experiment is run.
In this sense the excitation is \emph{designed} rather than assumed, and the dynamics enters only through the requirement that the saturation regions are reachable.
The paper turns this idea into three results.

\paragraph{A general identification principle}
An excitation design is a finite list of saturation values $v_1,\ldots,v_N$ spanning $\R^n$, together with disjoint time intervals $J_1,\ldots,J_N$.
A trajectory realizes the design if $S(x(t))=v_i$ on each $J_i$ (Definition~\ref{def: SIE}).
Proposition~\ref{prop:main-identification} shows that along any such trajectory an explicit closed-form estimator returns $W$ exactly.
This principles naturally doesn't depend on the dynamics $f$; neither does it require an observer. It is also general in terms of rank of $W$.

\paragraph{Excitation by open-loop controls}
We then give conditions under which a design can be realized.
Namely, we consider dynamics of the form
\[
    \dot x=Ax+\phi(x,u,t)+Bu,
\]
where $A$ is a perturbation of a Hurwitz matrix $A_0$ and $\phi$ an unknown bounded perturbation.
This class contains the Wilson--Cowan networks mentioned above.
The saturation regions are assumed to contain coradiant sets (i.e. stable by dilation of factor larger than $1$), as they do for componentwise saturations (see Definition~\ref{def:persistent-saturation}).
We assume that the image of $A_0^{-1}B$ intersects each saturation region, which holds for generic $A_0$ for componentwise saturations as soon as $\rank B\ge 2$ (see Proposition~\ref{prop:rank-B}).
Under this condition, Theorems~\ref{thm:perturbed-A} and~\ref{thm: arbitrary time} provide explicit controls, computed from the known data alone, such that every solution starting in a prescribed ball realizes an excitation design.
The controls are uniform with respect to the initial condition and to the unknown perturbation, within known bounds.
The first theorem only needs the linear part up to a small perturbation, and the second one gives the cases in which the design can be realized in arbitrarily short time.
Based on the previous identification principle, these theorems construct an open-loop experiment that identifies the matrix $W$, under mild structural assumptions on the dynamics and the saturations.
This is in a sense the main result of the paper.
We then pursue with reconstruction of the state and additional parameters, under significantly stronger assumptions.

\paragraph{Output feedback tracking}
Once $W$ is known, the problem can be reversed.
Instead of driving the state into the saturation regions, the control may be used to keep it away from them, in the region where the output is an invertible function of the state, if the saturations are of a more special ---but usual--- kind, and if the matrix $B$ is invertible (see Section~\ref{subsec: state}).
Under it, Theorem~\ref{thm: output feedback} constructs an output feedback, with a sufficiently large gain, that brings every trajectory into the unsaturated region in prescribed time and tracks a reference curve there, so that the state is recovered by componentwise inversion of the saturation.

Together, the three results give a three-stage procedure, stated in Section~\ref{subsec: full identification}:
An open-loop experiment identifies $W$, an output feedback gives access to the state, and, for Wilson--Cowan networks, a reference chosen to excite the dynamics yields the coupling matrix by a second closed-form regression (Theorem~\ref{thm: full identification}).

\begin{remark}
    This paper extends \cite{annabi2025identification}, where we considered a layered network of Wilson--Cowan type in which a single population is fully measured, and the interconnection weights were identified layer by layer by an online recursive procedure.
    Here, on the other hand, the dynamics is unknown apart from its linear part, the considered saturations are more general than the classical componentwise one, and identification relies on a single open-loop experiment and a closed-form estimator.
\end{remark}

\paragraph{Structure of the paper}
Section~\ref{sec: main results} presents the main results.
Section~\ref{sec: proofs} contains the geometric lemmas on coradiant sets, the construction of the controls and the proofs of the identification theorems.
Section~\ref{sec: feedback proofs} is devoted to the  tracking by output feedback, which yields the state, including the effect of measurement noise.
Section~\ref{sec: simulation} illustrates the identification procedure on a two-dimensional network with noise.

\subsection*{Notation}
We let $\llb a,b\rrb$ be the set of integers between $a$ and $b$,  $\mathring K$ be the topological interior of a set $K$,
and $\mathcal{B}(v,\rho)$ be the closed ball of radius $\rho$ centered at $v\in\R^n$.
We denote by $\|\cdot \|$ both the Euclidean norm on $\R^n$ and the associated operator norm (spectral norm) for matrices.
We also denote the characteristic function of a set $K$ by $\mathbf{1}_K$, and the Moore--Penrose pseudoinverse of a matrix $M$ by $M^+$.

\section{Main results} \label{sec: main results}

Our results, as stated in this section, get progressively stronger or more precise under progressively more restrictive assumptions on the dynamics ($f$) and the saturation function $S$ in system \eqref{eq: intro model}.
As already mentioned, our main case of interest is that of Wilson--Cowan networks, which involve a more particular form of $f$ (see Section~\ref{subsec: full identification}), and functions $S$ that are componentwise saturation in the following sense.
\begin{definition}
    \label{def:componentwise-saturation}
        A continuous function $S:\R^n\to \R^n$ is a \emph{componentwise saturation} if $S(x)=(S_0(x_1),\ldots, S_0(x_n))^\top$ with $S_0:\R\to \R$ nondecreasing and such that, for some $\ubar R < \bar R$, $S_0$ is strictly increasing on $[\ubar R,\bar R]$ and
    \[
        S_0(\xi) = 0 \qquad \text{if }\xi\le \ubar R,
        \qquad\text{and}\qquad 
        S_0(\xi) = 1 \qquad \text{if }\xi\ge \bar R.
    \]
\end{definition}
Although our assumptions always include Wilson--Cowan networks, we will, consider more general saturation types than above (see Definitions~\ref{def: saturation} and \ref{def:persistent-saturation}) 
highlighting the structural features underlying our results. 

\subsection{A general identification result} \label{subsec: identification}

We first state a preliminary identification result that holds for system \eqref{eq: intro model} in a fairly general setting. 
Concerning the vector field $f$, we only need it to be (for instance) continuous w.r.t. $(x,u)$ and measurable w.r.t. $t$, so that solutions on time interval $[0,T]$, to be taken in the sense of Carathéodory \cite{filippov1988differential}, make sense in Definition~\ref{def: SIE} and Proposition~\ref{prop:main-identification}.

For the immediate discussion, we only require the following broad definition of saturation.
\begin{definition}
    \label{def: saturation}
    A continuous function $S:\R^n\to\R^n$ is a \emph{saturation function} if there exist a family of closed sets with nonempty interior $K_1,\ldots, K_N\subset\R^n$ and vectors $v_1,\ldots, v_N\in\R^n$, such that
    \begin{equation}
        S(x) = v_i , \qquad \forall x\in K_i,\ \forall i\in \llb1,N\rrb .
    \end{equation}
    Each set $K_i$ is a \emph{saturation region} of $S$ and $v_i$ is its \emph{saturation value}.
\end{definition}
A componentwise saturation in the sense of Definition~\ref{def:componentwise-saturation} is naturally a saturation function: just take $I=\{0,1\}^n$, $v_i=i,\;i\in I$, and, with $R=\max\{|\ubar R|, |\bar R|\}$,
\begin{multline}
    \label{eq: componentwise cone}
        K_v := \{x\in \mathcal O_v\mid |x_j|\ge R, j\in \llb 1,n\rrb\}, 
        \\
        \qquad\text{with}\ \mathcal O_v := \{x\in \R^n\mid x_j\ge 0 \text{ if } v_j=1, \, x_j\le 0 \text{ if } v_j =0 \}
\end{multline}
(i.e. $\mathcal O_v$ is the closed orthant carrying the pattern $v$).

When the structure of the saturation is sufficiently rich, i.e., when the saturation values span $\R^n$, the unknown matrix $W$ can be recovered exactly from the output of a trajectory that visits the saturation regions in a prescribed order and for prescribed durations.
This is codified in the following definition, which is the main assumption of the identification result.

\begin{definition}
    \label{def: SIE}
    An \emph{excitation design} for the saturation function $S$ is a collection $\mathcal E = \{ (v_i,J_i)\mid i\in I \}$, where $I\subset \llb 1,N\rrb$ is a finite index set which up to relabeling we take as $I=\llb 1,M\rrb$, the $v_i$ are saturation values of $S$,
    and the $J_i\subset \R$ are pairwise disjoint closed intervals of positive length. Finally, we require that 
    \begin{equation} \label{eq: rank condition}
        \rank\,[v_{1},\ldots,v_{M}] = n.
    \end{equation}

    An absolutely continuous trajectory $x:[0,T]\to\R^n$ is a \emph{realization of $\mathcal{E}$ in time $T>0$} if $J_i\subset[0,T]$ for every $i$ and
    \begin{equation} \label{eq: star}
        S(x(t))=v_{i} \qquad \text{for all } t\in J_i \text{ and all } i\in\llb 1,M\rrb .
    \end{equation}
    We say that $x$ is \emph{saturation-exciting} if it realizes an excitation design.

    Finally, the excitation design $\mathcal{E}$ is said to be \emph{uniformly realizable in time $T>0$ and radius $\rho>0$} if there exists a control $u:[0,T]\to\R^m$ such that any solution of \eqref{eq: intro model} under this control $u$ with initial condition $\|x(0)\| \leq \rho$ is defined on $[0,T]$ and is a realization of $\mathcal{E}$ in time $T$.
\end{definition}

Section~\ref{subsec: structural} provides sufficient conditions for the existence of a uniformly realizable excitation design.

Given an excitation design $\mathcal{E}$, write $\tau_i:=|J_i|>0$ for the length of the $i$-th interval and define the excitation Gramian and the estimator
\begin{equation} \label{eq: estimator}
    \Gamma(\mathcal{E}):=\sum_{i=1}^M\tau_i\, v_iv_i^\top\in\R^{n\times n},
    \quad
    E(y;\mathcal{E}):=\left(\sum_{i=1}^M\int_{J_i}y(s)\,v_i^\top\,ds\right)\Gamma(\mathcal{E})^{-1}\in\R^{q\times n}.
\end{equation}

The following observation is at the core of our identification procedure.
\begin{proposition}\label{prop:main-identification}
    Let $\mathcal{E}$ be an excitation design and let $(x,u,y)$ be a solution of \eqref{eq: intro model} on $[0,T]$ whose state realizes $\mathcal{E}$.
    Then $\Gamma(\mathcal{E})$ is invertible and
    \begin{equation}
        E(y;\mathcal{E})=W .
    \end{equation}
\end{proposition}

\begin{proof}
    The matrix $\Gamma(\mathcal{E})$ is symmetric and positive semidefinite.
    Indeed, for any $z\in\R^n$, we have $z^\top\Gamma(\mathcal{E})z=\sum_i\tau_i(v_i^\top z)^2\ge 0$. Moreover, $z^\top\Gamma(\mathcal{E})z=0$ forces $v_i^\top z=0$ for every $i$.
    By \eqref{eq: rank condition}, this implies that $z=0$.
    To conclude, observe that by \eqref{eq: star} the output verifies $y(t)=WS(x(t))=Wv_i$ for every $t\in J_i$.
    Therefore
    \begin{equation}
        \sum_{i=1}^N\int_{J_i}y(s)v_i^\top\,ds=\sum_{i=1}^N\tau_i\,Wv_iv_i^\top=W\,\Gamma(\mathcal{E}).
    \end{equation}
    Multiplying on the right by $\Gamma(\mathcal{E})^{-1}$ yields the result.
\end{proof}

A recursive least-squares implementation of this regression is possible, see \cite{goodwin1984adaptive} or \cite[Sec.\ 7.3]{ljung1999system}.

\begin{remark}
    \label{rem: partial rank}
    The hypotheses of Proposition~\ref{prop:main-identification} can be relaxed in two ways.
    First, suppose the values of the design span only a proper subspace $V:=\spann\{v_1,\ldots,v_M\}$ of $\R^n$, so that \eqref{eq: rank condition} fails.
    The Gramian $\Gamma(\mathcal{E})$ is then singular with range $V$, but the data still determine part of $W$.
    Indeed, replacing $\Gamma(\mathcal{E})^{-1}$ with the Moore--Penrose pseudoinverse in the definition of $E(y;\mathcal{E})$, the above argument yields
    \begin{equation}
        E(y;\mathcal{E})=W P_V,
    \end{equation}
    where $P_V$ is the orthogonal projector onto $V$.
    Second, the equality of the source and target dimensions of $S$ is not used in the proof.
    If $S:\R^n\to\R^{d}$ and $W\in\R^{q\times d}$, it suffices that the saturation values of the visited regions span $\R^{d}$.
\end{remark}

\begin{remark}[Measurement noise]
    \label{rem: noise identification}
    Let the output be measured as $\tilde y=y+\nu$ along a trajectory realizing $\mathcal{E}$.
    For open-loop controls (such as those of Section~\ref{subsec: structural}), {a bounded} noise $\nu$ does not affect the realization, and the linearity of \eqref{eq: estimator} then yields
    \[
        \|E(\tilde y;\mathcal{E})-W\|\le\|\nu\|_\infty\bigl(\sum_{i=1}^M\tau_i\|v_i\|\bigr)\|\Gamma(\mathcal{E})^{-1}\|
        \le \|\nu\|_\infty\bigl(\sum_{i=1}^M\tau_i\|v_i\|\bigr)\frac{\|(\mathcal V\mathcal V^\top)^{-1}\|}{ \min_i\tau_i} ,
    \]
    where we let $\mathcal V:=[v_1,\ldots,v_M]$.
\end{remark}

\subsection{Realizing excitation designs}
\label{subsec: structural}

Proposition~\ref{prop:main-identification} is of use only if an excitation design can be produced.
Here, we consider the following specialization of \eqref{eq: intro model},
\begin{equation} \label{eq: system}
    \begin{cases}
        \dot x = Ax + \phi(x,u,t) + Bu, \\
        y = WS(x).
    \end{cases}
\end{equation}
We assume that
\begin{itemize}
\item $A$ is a  (small) perturbation of a known Hurwitz matrix $A_0$, i.e., 
    \begin{equation}\label{eq:A0A1} A=A_0+A_1 \end{equation}
     with $A_1$ unknown and satisfying $\|A_1\|\le\bar a$ for some known $\bar a\ge0$,
    \item $\phi$ is an unknown continuous and bounded function, with $\|\phi\|_\infty\le\bar\phi$ for some known $\bar\phi>0$, and
    \item $B\in\R^{n\times m}$ is known.
\end{itemize}
This setting includes Wilson--Cowan networks with an unknown coupling matrix \cite{Wilson1972,coombes2018networks}, that we discuss in Section~\ref{subsec: full identification}.
As for the function $S$, we now on require it to be a coradiant saturation in the following sense.
\begin{definition}
    \label{def:persistent-saturation}
    The function $S$ is a \emph{coradiant saturation} if there exist $n$ closed coradiant sets\footnote{Recall that a set $K\subset \R^n$ is coradiant if $\lambda x\in K$ for all $x\in K$ and $\lambda\ge 1$.} $K_1,\ldots,K_{n} \subset \R^{n}$ with non-empty interior, and linearly independent vectors $v_1,\ldots,v_{n} \in \R^{n}$ such that
    \begin{equation} \label{eq: saturation on cones}
        S(x) = v_i, \quad  \forall x \in K_i, \quad i \in \llb 1, n\rrb.
    \end{equation}
\end{definition}

Coradiant saturations are clearly saturations in the sense of Definition~\ref{def: saturation}.
Componentwise saturations (Definition~\ref{eq: componentwise cone}) are coradiant saturations because the sets $K_v$ in \eqref{eq: componentwise cone} are coradiant, and $\{0,1\}^n$ contains the canonical basis of $\R^n$.

 \smallskip
 
The following result is proved in Section~\ref{subsec: proofs theorems}, based on the beginning of Section~\ref{sec: proofs}.

\begin{theorem}
    \label{thm:perturbed-A}
    Let $S$ be a coradiant saturation, and assume that
    \begin{equation}
        \label{eq: condition A_0}
        \Ima(A_0^{-1}B) \cap \mathring{K}_i \neq \varnothing \qquad \forall i \in \llb 1, n\rrb.
    \end{equation}
    Then, there exists $\varepsilon_0>0$ with the following property:
    for any $\bar a<\varepsilon_0$, there exists $T^*\ge0$ such that for every $\rho>0$ and every $T>T^*$ there exists an excitation design $\mathcal{E}$ which is uniformly realizable in time $T$ and radius $\rho$ for all $A_1$ with $\|A_1\|\le\bar a$ and all $\phi$ with $\|\phi\|_\infty\le\bar\phi$, via a control that does not depend on them.
    Moreover, there is $\bar\rho\ge\rho$, independent of $A_1$ and $\phi$, such that these solutions satisfy $\|x(t)\|\le\bar\rho$ for every $t\in[0,T]$.
\end{theorem}

\begin{remark}
    In the above result, $\varepsilon_0$ depends only on $A_0$, $B$ and $K_1,\ldots,K_n$, while $T^*$ depends also on $\bar a$ and is given by \eqref{eq: T star}.
    The control and the radius $\bar\rho$ are explicit. The control is piecewise constant, and both depend on the known data alone, namely $A_0$, $B$, $K_1,\ldots,K_n$, $\bar a$, $\bar\phi$, $\rho$ and $T$.
\end{remark}

Condition~\eqref{eq: condition A_0} is always satisfied if $\rank B\ge n$ and never if $\rank B=1$ and $n\ge3$.
Depending on the saturation, this condition can be weakened: for componentwise saturations we have the following proposition, proved at the end of Section \ref{subsec: proofs theorems}. 
\begin{proposition}\label{prop:rank-B}
        Let $S$ be a componentwise saturation.       
        Then, for any $B$ with $\rank B\ge 2$, the set of Hurwitz matrices $A_0$ such that condition \eqref{eq: condition A_0} holds for suitable sets $K_{\mathbf s}$ of \eqref{eq: componentwise cone} is open and dense in the set of Hurwitz matrices.
\end{proposition}

The idea underlying the construction in Theorem~\ref{thm:perturbed-A} is the fact that, for a linear system $\dot x=A_0x+Bu$ with $A_0$ Hurwitz, for any $b\in\Ima(B)$ one can steer the system towards $-\mu A_0^{-1}b$ using the constant controls $Bu\equiv\mu b$.
In particular, under \eqref{eq: condition A_0}, taking $\mu$ sufficiently large allows to sequentially drive the state to the desired regions $\mathring{K}_1,\ldots,\mathring{K}_n$.
Making this procedure insensitive to the perturbation $A_1$ requires then that it be sufficiently small, while the bounded term $\phi$ is dominated by taking $\mu$ large.

Under additional assumptions on the unperturbed part $A_0$, the design can be realized in arbitrarily short time.

\begin{theorem} \label{thm: arbitrary time}
    Under the assumptions of Theorem~\ref{thm:perturbed-A} hold. assume that
    \begin{itemize}
        \item[i.] either $\bar a=0$, that is, $A=A_0$ is known, and the pair $(A_0,B)$ is controllable,
        \item[ii.] or $A_0=-\lambda\idty$ for some $\lambda>0$.
    \end{itemize}
    Then, the conclusion of Theorem~\ref{thm:perturbed-A} holds with $T^*=0$, that is, for every $T>0$.
\end{theorem}

\begin{remark}
    Also in this case the control and the radius $\bar\rho$ are explicit and depend on the known data alone, namely $A_0$ (or $\lambda$), $B$, $K_1,\ldots,K_n$, $\bar a$, $\bar\phi$, $\rho$ and $T$.
    The control is piecewise constant in case \emph{ii.}, while in case \emph{i.} each step uses control \eqref{eq: gramian control}.
\end{remark}

\subsection{Tracking by output feedback and state recovery} \label{subsec: state}

We now assume that the matrix $W$ is known, e.g.\ by implementing Proposition~\ref{prop:main-identification} and Theorem \ref{thm:perturbed-A}, and proceed with state reconstruction,
under three more assumptions.
First, we suppose that $B$ in \eqref{eq: system} is the identity matrix\footnote{If $B$ is invertible, we can always reduce to this case by a change of variables $u\mapsto Bu$.}, and consider the following further specialization of \eqref{eq: system}:
    \begin{equation} \label{eq: system-B-invertible}
        \begin{cases}
            \dot x = Ax + \phi(x,u,t) + u, \\
            y = WS(x).
        \end{cases}
    \end{equation}
Second, we assume that \emph{$S$ is a componentwise saturation}\footnote{
   If, instead, $S=PS_1$ with $P$ known and invertible and $S_1$ a componentwise saturation, then the second line in \eqref{eq: system-B-invertible} reads  $y=(WP)S_1(x)$, and everything below applies with $WP$ in place of $W$.
}
rather than merely a coradiant saturation; this adds invertibilty properties outside saturation regions: $S$ is injective on $\Omega:=[\ubar R,\bar R]^{n}$.
Third, we assume that the \emph{known matrix $W$ is left invertible} (i.e., $\rank W=n$; this requires $q\ge n$); then, fixing a left inverse $H$ of $W$, the second line in \eqref{eq: system-B-invertible} reads
    \begin{equation} \label{eq: componentwise output}
        Hy=S(x) .
    \end{equation}
The following result, proved in Section~\ref{sec: feedback proofs}, allows tracking of any reference $\gamma$ staying in $\mathring\Omega$, via an explicit output feedback law; 
since $S$ is invertible on $\mathring\Omega$, the state can then be read off the output from \eqref{eq: componentwise output}.

\begin{theorem} \label{thm: output feedback}
    Let the previous assumptions hold.
    Let $\rho>0$ and $\varepsilon>0$, let $0<T_1<T_2$, and let $\gamma$ be a $C^1$ curve with bounded derivative taking its values in a compact subset $\mathcal{K}$ of $\mathring\Omega$.
    Then there exist $k^\star>0$ and continuous maps $\eta_k:\R^q\times\R^n\to\R^m$, $k>0$, such that, for every $k>k^\star$, every solution of \eqref{eq: system-B-invertible}, with $\|x(0)\|\le\rho$, under the output feedback $u(t)=\eta_k(y(t),\gamma(t))$ is defined for all positive times
    and satisfies
    \begin{equation} \label{eq: feedback conclusion}
        x(t)\in\mathring\Omega \quad \forall t\ge T_1
        \qquad\text{and}\qquad
        \|x(t)-\gamma(t)\|\le\varepsilon \quad \forall t\ge T_2 ,
    \end{equation}
    and $x_j(t)=S_0^{-1}\bigl((Hy(t))_j\bigr)$ for $t\ge T_1$ and $j\in\llb1,n\rrb$, where $S_0^{-1}:[0,1]\to [\ubar R,\bar R]$ is the inverse of $S_0|_{[\ubar R,\bar R]}$.

    Moreover, if $S_0^{-1}$ is locally Lipschitz on $(0,1)$, there exist $\delta_0,c_\nu>0$ such that for every measurable $\nu:[0,+\infty)\to\R^q$ with $\|\nu(t)\|_\infty\le\delta\le\delta_0$, and the 
    output feedback $u(t)=\eta_k(y(t)+\nu(t),\gamma(t))$, the same conclusions as above hold with $\varepsilon+c_\nu\delta$ in place of $\varepsilon$ in \eqref{eq: feedback conclusion}. Also, the recovered state $\hat x_j(t):=S_0^{-1}\bigl((H(y(t)+\nu(t)))_j\bigr)$ is well defined for $t\ge T_1$ and satisfies $\|\hat x(t)-x(t)\|_\infty\le c_\nu\delta$.
\end{theorem}

\begin{remark}
    In the above result, the maps $\eta_k$ and the constants $k^\star$, $\delta_0$ and $c_\nu$ are explicit and given in Section~\ref{sec: feedback proofs}.
    In particular, the maps $\eta_k$ depend only on $W$, $S$, $S_0$ and $\mathcal{K}$, while $k^\star$, which is defined in \eqref{eq: k star}, does not depend on $W$.
    None depends on $\nu$, and $c_\nu$ does not depend on $k$, so the floor $c_\nu\delta$ is not lowered by a larger gain.
\end{remark}

\subsection{Identification of the full system} \label{subsec: full identification}
We conclude this section by showing how to combine the identification of $W$ and the output feedback of Theorem~\ref{thm: output feedback} to recover the state, and then the dynamics, from the output.
We showcase this procedure in the case of Wilson--Cowan networks \cite{Wilson1972,coombes2018networks},
where the nonlinear part of the dynamics \eqref{eq: system-B-invertible} is given by 
\begin{equation}
    \label{eq:Csigma}
    \phi(x,u,t)=C\sigma(x),
\end{equation}
with $\sigma$ a known continuous bounded function  (for instance, a saturation, unrelated to $S$) and $C$ an unknown constant matrix.

\begin{theorem} \label{thm: full identification}
    Consider \eqref{eq: system-B-invertible} with $S$ a componentwise saturation and $W$ injective, and assume that $\bar a<\varepsilon_0$, where $\varepsilon_0$ and $T^*$ are given by Theorem~\ref{thm:perturbed-A}.
    Let $\rho>0$, $T>T^*$, $\varepsilon>0$ and $0<T_1<T_2$, and let $\gamma$ and $\mathcal K$ be as in Theorem~\ref{thm: output feedback}.
    Then there exist an excitation design $\mathcal E$, a constant $k^\star>0$ and, for every $k>k^\star$, a control $u$, open loop on $[0,T)$ and an output feedback on $[T,+\infty)$, both independent of $A_1$, $\phi$,  and without prior knowledge of $W$, such that every solution with $\|x(0)\|\le\rho$ satisfies the following.
    \begin{itemize}
        \item[i.] $\hat W:=E(y|_{[0,T]};\mathcal E)=W$.
        \item[ii.] $x(t)\in\mathring\Omega$ and $x_j(t)=S_0^{-1}\bigl((\hat W^+y(t))_j\bigr)$ for $t\ge T+T_1$ and $j\in\llb1,n\rrb$, and $\|x(t)-\gamma(t-T)\|\le\varepsilon$ for $t\ge T+T_2$.
        \item[iii.] Assume moreover $A=A_0$ is known and $\phi(x,u,t)=C\sigma(x)$, with $C\in\R^{n\times n}$ unknown and $\sigma:\R^n\to\R^n$ known, continuous and bounded.
        
        Assume there exist $ p^1,\ldots,p^n\in\mathcal K$ such that $P:=[\sigma(p^1),\ldots,\sigma(p^n)]$ is invertible.
        Pick $\gamma$ such that $\gamma(t)=  p^\ell$ on disjoint intervals $t\in I_\ell=[a_\ell,b_\ell]\subset [T_2,\infty)$, $a_\ell<b_\ell$, for all $\ell\in\llb1,n\rrb$, and let
        \begin{equation} \label{eq: C regression}
            d_\ell:=x(T+b_\ell)-x(T+a_\ell)-\int_{T+I_\ell}(Ax+u)\,dt ,
            \quad
            \zeta_\ell:=\int_{T+I_\ell}\sigma(x)\,dt .
        \end{equation}
        Then there exists $\varepsilon_1>0$, depending only on $\sigma$ and $p^1,\ldots,p^n$, such that if $\varepsilon<\varepsilon_1$, the matrix $Z:=[\zeta_1,\ldots,\zeta_n]$ is invertible and $C=[d_1,\ldots,d_n]\,Z^{-1}$.
    \end{itemize}
\end{theorem}

\begin{proof}
    The proof is divided into three parts, corresponding to the three items of the statement.

    \emph{Step 1: Identification of $W$.}
    As observed in Section~\ref{subsec: structural}, $S$ is a coradiant saturation with the sets $K_{e_j}$ of \eqref{eq: componentwise cone} and $v_j=e_j$.
    Since $B=\idty$, condition \eqref{eq: condition A_0} holds.
    Hence Theorem~\ref{thm:perturbed-A} gives a control $u_{\mathrm{ol}}$ on $[0,T]$, a design $\mathcal{E}$ and a radius $\bar\rho$, independent of $A_1$ and $\phi$, such that every solution with $\|x(0)\|\le\rho$ realizes $\mathcal E$ and satisfies $\|x(T)\|\le\bar\rho$.
    Proposition~\ref{prop:main-identification} then gives item \emph{i.}

    \emph{Step 2: Tracking.}
    Let $\hat\eta_k$ be the feedback of Theorem~\ref{thm: output feedback} built with $\hat W$ and the left inverse $H=\hat W^+$, and consider
    \begin{equation} \label{eq: two-stage control}
        u(t)=
        \begin{cases}
            u_{\mathrm{ol}}(t) ,                     & t\in[0,T) , \\
            \hat\eta_k\bigl(y(t),\gamma(t-T)\bigr) , & t\ge T .
        \end{cases}
    \end{equation}
    Since $x(T+\cdot)$ solves \eqref{eq: system-B-invertible} with $\phi$ shifted in time, which obeys the same bound $\bar\phi$, and $\|A\|\le\|A_0\|+\bar a$ is known, the conclusions of Theorem~\ref{thm: output feedback}, applied with $\bar\rho$ in place of $\rho$, hold for $k>k^\star$ with $t$ replaced by $t-T$.
    This is item \emph{ii.}

    \emph{Step 3: Identification of $C$.}
    By item \emph{ii.}, $x$ is known on $T+I_\ell\subset[T+T_1,+\infty)$, so $d_\ell$ and $\zeta_\ell$ are computed from $y$ and $u$.
    Integrating the dynamics $\dot x = Ax + C\sigma(x)+u$ over $T+I_\ell$ gives $d_\ell=C\zeta_\ell$, so it suffices to show that $Z$ is invertible.
    Let $\omega_\sigma$ be a modulus of continuity of $\sigma$ on $\Omega$.
    Since $\|x(T+t)- p^\ell\|\le\varepsilon$ for $t\in I_\ell$, we have $\|\zeta_\ell-|I_\ell|\,\sigma( p^\ell)\|\le|I_\ell|\,\omega_\sigma(\varepsilon)$.
    Thus, $Z=(P+R)\,\mathrm{diag}(|I_1|,\ldots,|I_n|)$ with $\|R\|\le\sqrt n\,\omega_\sigma(\varepsilon)$.
    In particular, $Z$ is invertible as soon as $\sqrt n\,\omega_\sigma(\varepsilon)<\|P^{-1}\|^{-1}$, which defines $\varepsilon_1$.
\end{proof}

Other identification procedures, such as those based on persistent excitation, can also be used at this stage. 
Accordingly, this procedure can be extended to more general $\phi$, as soon as the corresponding dynamics are identifiable by prescribing the state and knowledge of the input.

With the noisy output of Theorem~\ref{thm: output feedback}, the estimator of item \emph{iii.} is evaluated on the recovered state $\hat x$, which is within $c_\nu\delta$ of $x$, while $u$ is known exactly, since it is the applied control.
Hence, if $\sigma$ is Lipschitz and $\varepsilon+c_\nu\delta<\varepsilon_1$, the error on $C$ is of order $\delta$.
More windows can also be combined by least squares, as in Section~\ref{sec: simulation}.

\begin{remark}
    If $A$ is also unknown, the same argument applies to $[A\ C]$ with the regressor $(x,\sigma(x))\in\R^{2n}$ and $2n$ points $ p^\ell$.
    Such points exist if and only if the components of $x$ and $\sigma(x)$ are linearly independent functions on $\mathring\Omega$, which holds for a componentwise $\sigma$ if and only if its profile $\sigma_0$ is not affine on $[\ubar R,\bar R]$.
\end{remark}

\section{Realization of the excitation designs} \label{sec: proofs}

In this section we prove Theorems~\ref{thm:perturbed-A} and~\ref{thm: arbitrary time}, which assume the saturation to be coradiant, as per Definition~\ref{def:persistent-saturation}. 
Hence, the natural choice for the vectors of the design $\mathcal E$ is $v_1,\ldots,v_n$.
Realizing such a design consists in bringing the state into $K_1$ and holding it there during the interval $J_1$, then into $K_2$ during $J_2$, and so, on up to $K_n$.
Thus, we need to do two things: \emph{(i)} construct a control bringing the state into one prescribed region and keeping it there, and \emph{(ii)} chain $n$ such controls.

Section~\ref{subsec: steering} addresses the first ingredient.
It measures how much room a coradiant region leaves around a given direction, and then solves the one-region problem. 
Section~\ref{subsec: concatenation} chains the steps, and Section~\ref{subsec: proofs theorems} proves the theorems.

\subsection{Steering into one region} \label{subsec: steering}

We introduce the following notion.
\begin{definition}
    \label{def: steering}
    Let $K\subset\R^n$ be closed and coradiant with nonempty interior and let $T>0$.
    We say that \eqref{eq: system} \emph{can be steered into $K$ in time $T$} if for every $\rho>0$ and every initial time $t_0\ge0$ there exist a bounded measurable control $u:[t_0,+\infty)\to\R^m$ and a radius $\rho'\ge\rho$ such that every solution of \eqref{eq: system} with $\|x(t_0)\|\le\rho$ satisfies
    \begin{equation} \label{eq: steering conclusion}
        x(t)\in K \quad \forall t\ge t_0+T ,
        \qquad
        \|x(t)\|\le\rho' \quad \forall t\ge t_0 .
    \end{equation}
\end{definition}

Observe that the same $T$ must work for every initial ball, but the control and the radius $\rho'$ may depend on $\rho$.
For coradiant saturations and for $K=K_i$, \eqref{eq: steering conclusion} implies $S(x(t))=v_i$ for every $t\ge t_0+T$.

In this section we present three lemmas, each of which establishes that \eqref{eq: system} can be steered into a coradiant region in time $T$, strictly positive for the first one and arbitrarily small for the second and third.
To state them, we need to introduce a measure of how much room a coradiant region leaves around a given direction.

\subsubsection{Margin of a coradiant region} \label{subsubsec: margin}

\begin{definition}
    \label{def: margin}
    Let $K\subset\R^n$ be closed and coradiant with nonempty interior, and let $v\in\mathring{K}$, $v\neq0$.
    The \emph{margin} of $v$ in $K$ is
    \begin{equation}
        \delta_K(v) := \min\left\{1,\ \frac{\dist(v,\R^n\setminus K)}{\|v\|}\right\}\in(0,1] .
    \end{equation}
\end{definition}

The margin measures the distance from $v$ to the boundary of $K$, relative to $\|v\|$.
On a convex cone it is the sine of the angle from $v$ to the nearest face.
On an orthant it is $\min_j|v_j|/\|v\|$.

The steering lemmas of Section~\ref{subsec: steering} are stated in terms of the following quantity, which depends only on the data of the problem, and not on a particular direction.

\begin{definition}
    Let $A'$ be Hurwitz and let $K\subset\R^n$ be closed and coradiant with nonempty interior.
    The \emph{best margin} of $K$ with respect to $A'$ is
    \begin{equation}
        \label{eq: best depth}
        \delta^*_K(A'):=\sup\bigl\{\delta_K(w):\ w\in\Ima(A'^{-1}B)\cap\mathring K\bigr\}\ \in\ [0,1] ,
    \end{equation}
    with the convention $\sup\varnothing=0$.
\end{definition}
Observe that $\delta^*_K(A')>0$ is equivalent to $\Ima(A'^{-1}B)\cap\mathring K\ne\varnothing$, and that every margin below $\delta^*_K(A')$ is exceeded by the margin of some $w\in\Ima(A'^{-1}B)\cap\mathring K$.

\begin{remark}
    The supremum in \eqref{eq: best depth} is in general not a maximum, and closedness of $K$ does not help.
    For the componentwise saturations the margin increases strictly along each ray, toward the limit $\min_j|w_j|/\|w\|$, which is therefore never attained.
\end{remark}

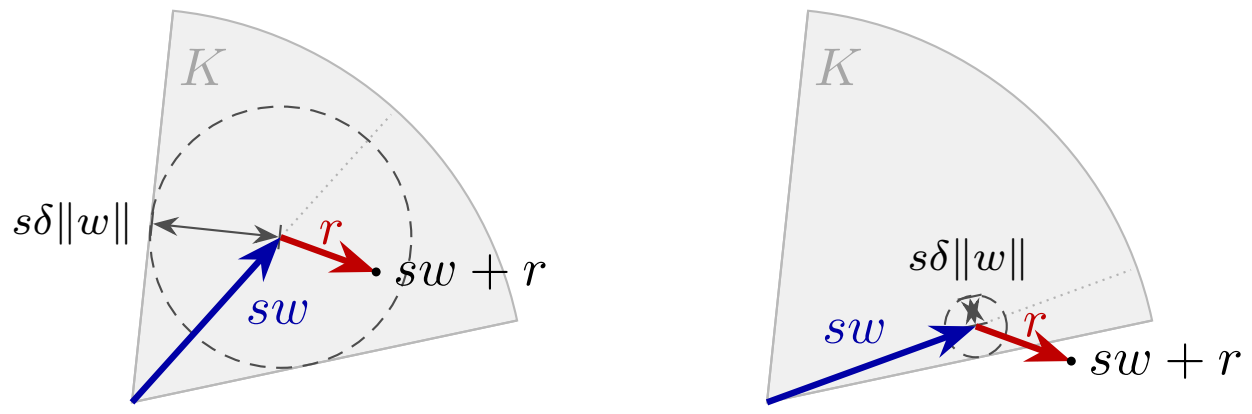
\begin{figure}
    \centering
    \begin{tikzpicture}[scale=0.42,>=Stealth,
            region/.style={fill=gray!12,draw=gray!55},
            room/.style={draw=black!70,densely dashed,thin},
            design/.style={->,very thick,blue!65!black},
            resid/.style={->,very thick,red!75!black}]

        \begin{scope}
            \fill[region] (0,0) -- (12:6.2) arc[start angle=12,end angle=84,radius=6.2] -- cycle;
            \draw[gray!55] (0,0) -- (12:6.2);
            \draw[gray!55] (0,0) -- (84:6.2);
            \node[gray!70,font=\small] at (78:5.4) {$K$};

            \draw[densely dotted,gray!60] (0,0) -- (48:6.1);
            \coordinate (c) at (48:3.5);
            \draw[room] (c) circle (2.057);

            \draw[|<->|,thin,black!70] (c) -- ++(174:2.057);
            \node[font=\scriptsize,anchor=east,inner sep=2pt] at (0.20,2.816) {$s\delta\|w\|$};

            \draw[design] (0,0) -- (c);
            \node[blue!65!black,font=\small,below right,inner sep=2pt] at (48:2.4) {$sw$};

            \coordinate (x) at ($(c)+(-20:1.6)$);
            \draw[resid] (c) -- (x);
            \node[red!75!black,font=\small,above,inner sep=2pt] at ($(c)+(-20:0.85)$) {$r$};
            \fill (x) circle (2pt);
            \node[font=\small,right,inner sep=3pt] at (x) {$sw+r$};

        \end{scope}

        \begin{scope}[xshift=10cm]
            \fill[region] (0,0) -- (12:6.2) arc[start angle=12,end angle=84,radius=6.2] -- cycle;
            \draw[gray!55] (0,0) -- (12:6.2);
            \draw[gray!55] (0,0) -- (84:6.2);
            \node[gray!70,font=\small] at (78:5.4) {$K$};

            \draw[densely dotted,gray!60] (0,0) -- (20:6.1);
            \coordinate (d) at (20:3.5);
            \draw[room] (d) circle (0.487);
            \draw[|<->|,thin,black!70] (d) -- ++(102:0.487);
            \node[font=\scriptsize,inner sep=2pt,anchor=south] at ($(d)+(102:0.55)$) {$s\delta\|w\|$};

            \draw[design] (0,0) -- (d);
            \node[blue!65!black,font=\small,above left,inner sep=2pt] at (20:2.2) {$sw$};

            \coordinate (y) at ($(d)+(-20:1.6)$);
            \draw[resid] (d) -- (y);
            \node[red!75!black,font=\small,above,inner sep=2pt] at ($(d)+(-20:1.0)$) {$r$};
            \fill (y) circle (2pt);
            \node[font=\small,right,inner sep=3pt] at (y) {$sw+r$};

        \end{scope}

    \end{tikzpicture}
    \caption{ Lemma~\ref{lemma: master}: $K$ contains the ball of radius $s\delta\|w\|$ centered at $sw$ (dashed).
        Left: the margin absorbs $r$.
        Right: with a small margin, the same $r$ leaves $K$.}
    \label{fig: margin}
\end{figure}

The following comparison is the core of the three steering lemmas below.
Figure~\ref{fig: margin} shows why the margin must exceed the residual.

\begin{lemma}
    \label{lemma: master}
    Let $K\subset\R^n$ be closed and coradiant with nonempty interior, let $w\in\mathring K$ with $w\neq0$, and let $\delta\in(0,\delta_K(w)]$.
    Let $s_0\ge1$, $c_1\ge0$ and $c_2\ge0$ satisfy
    \begin{equation} \label{eq: master condition}
        s_0\,(\delta-c_1)\,\|w\|\ \ge\ c_2 .
    \end{equation}
    Then, for every $s\ge s_0$ and every $r\in\R^n$ such that
    \begin{equation} \label{eq: master shape}
        \|r\|\ \le\ s\,c_1\|w\|+c_2 ,
    \end{equation}
    we have $sw+r\in K$.
\end{lemma}

\begin{proof}
    Since $c_2\ge0$ and $s_0\|w\|>0$, condition \eqref{eq: master condition} forces $c_1\le\delta$.
    Using $s\ge s_0$, \eqref{eq: master shape} and then \eqref{eq: master condition},
    \begin{equation} \label{eq: master gap}
        s\delta\|w\|-\|r\|
        \ \ge\ s(\delta-c_1)\|w\|-c_2
        \ \ge\ s_0(\delta-c_1)\|w\|-c_2
        \ \ge\ 0 .
    \end{equation}
    In particular $\|r\|\le s\delta\|w\|$.
    Denote by $\mathcal{B}(v,\rho)$ the closed ball of radius $\rho$ centered at $v$.
    Since $\delta\le\delta_K(w)$ and $K$ is closed, the definition of the margin gives $\mathcal{B}(w,\delta\|w\|)\subset K$.
    Since $s\ge1$ and $K$ is coradiant, it follows that
    \begin{equation*}
        \mathcal{B}(sw,s\delta\|w\|)\ =\ s\,\mathcal{B}(w,\delta\|w\|)\ \subset\ sK\ \subset\ K ,
    \end{equation*}
    and therefore $sw+r\in K$.
\end{proof}

\subsubsection{The steering lemmas}
For simplicity we present all statements and proofs with $t_0=0$, the general case being obtained by a time translation.
In the three lemmas below, the direction $b$ and the threshold $\mu_0$, hence the control, are computed from $A_0$, $B$, $K$, $T$, $\rho$, $\bar\phi$ and $\bar a$ only, and not from $A_1$ or $\phi$.

Since $A_0$ is Hurwitz, there exist constants $M_0\ge 1$ and $\alpha_0>0$ such that
\begin{equation} \label{eq: exp bound}
    \|e^{A_0t}\|\le M_0e^{-\alpha_0 t}, \qquad t\ge0.
\end{equation}
Once a solution $x$ and a control $u$ are fixed, we abbreviate $\phi(s):=\phi(x(s),u(s),s)$.
In particular, the solution $x$ is given by the variation of constants formula
\begin{equation} \label{eq: voc}
    x(t)=e^{At}x(0)+\int_0^te^{A(t-s)}\bigl[\phi(s)+Bu(s)\bigr]ds , \qquad t\ge0 .
\end{equation}

Our first steering lemma uses a constant control, which forces a lower bound on the time needed to steer the system into the coradiant region.
The control is built from $A_0$, while the equilibrium it reaches, $-A^{-1}b$, involves the unknown $A_1$.
Recall that a bound $\bar a\ge\|A_1\|$ is known, and write
\begin{equation} \label{eq: theta}
    p:=\frac{M_0}{\alpha_0}\,\bar a , \qquad \vartheta:=\frac{p}{1-p}\quad(p<1) , \qquad \alpha_1:=\alpha_0-M_0\bar a .
\end{equation}
Since $\|A_0^{-1}\|\le M_0/\alpha_0$ by \eqref{eq: exp bound}, we have $\|A_0^{-1}A_1\|\le p$, and $\vartheta$ bounds how far $A_1$ can move that equilibrium relative to $\|A_0^{-1}b\|$.

\begin{lemma}[Steering by a constant control]
    \label{lemma: steering perturbed}
    Let $A_0$ be Hurwitz with constants $M_0,\alpha_0$ as in \eqref{eq: exp bound}, and let $K\subset\R^n$ be closed and coradiant with nonempty interior.
    Set $\delta^*:=\delta^*_K(A_0)$ and assume that
    \begin{equation} \label{eq: A1 condition}
        \bar a\ <\ \frac{\alpha_0}{M_0}\cdot\frac{\delta^*}{1+\delta^*} .
    \end{equation}
    With $\vartheta$ and $\alpha_1$ as in \eqref{eq: theta}, set
    \begin{equation} \label{eq: T min perturbed}
        T_{\min}:=\frac{1}{\alpha_1}\ln\frac{M_0}{\hat\delta} ,
        \quad\text{where}\quad
        \hat\delta:=\frac{\delta^*-\vartheta}{1+\vartheta} .
    \end{equation}
    Then, \eqref{eq: system} can be steered into $K$ in every time $T>T_{\min}$.
    More precisely, given $T>T_{\min}$, there exists $b\in \Ima(B)$ such that, for every $\rho>0$, there exists $\mu_0\ge1$ such that, for every $\mu\ge\mu_0$,
    the constant control $Bu\equiv\mu b$ drives every solution of \eqref{eq: system} with $\|x(0)\|\le \rho$ so that \eqref{eq: steering conclusion} holds with $\rho'=R^0_\mu$ given by
    \begin{equation} \label{eq: state bound perturbed}
        R^0_\mu:=\mu(1+M_0)(1+\vartheta)\|A_0^{-1}b\|+M_0\rho+\frac{M_0\bar\phi}{\alpha_1} .
    \end{equation}
\end{lemma}

\begin{proof}
    We start by estimating the decay of $e^{At}$.
    Recall that
    \[
        e^{At}=e^{A_0t}+\int_0^te^{A_0(t-s)}A_1e^{As}ds.
    \]
    Hence, the bound \eqref{eq: exp bound} and $\|A_1\|\le\bar a$ show that $\psi(t)\le M_0+M_0\bar a\int_0^t\psi(s)ds$, for $\psi(t):=e^{\alpha_0t}\|e^{At}\|$.
    Grönwall's lemma gives
    \begin{equation} \label{eq: perturbed decay}
        \|e^{At}\|\ \le\ M_0e^{-\alpha_1t} , \qquad t\ge0 .
    \end{equation}
    By \eqref{eq: A1 condition}, $M_0\bar a<\alpha_0$, so $\alpha_1>0$, and in particular $A$ is Hurwitz.

    Let now $b\in\Ima(B)$ be such that $w_0:=-A_0^{-1}b\in\mathring K$ and
    \begin{equation} \label{eq: direction perturbed}
        M_0e^{-\alpha_1T}\ <\ \hat\delta_0:=\frac{\delta_0-\vartheta}{1+\vartheta} ,
        \qquad\text{where } \delta_0:=\delta_K(w_0) .
    \end{equation}
    Such a $b$ exists by definition of the best margin, since $\hat\delta_0\to\hat\delta$ as $\delta_0\to\delta^*$.
    In the following, we first determine the location of the equilibrium $w:=-A^{-1}b$ actually reached by the control with respect to the boundary of $K$.
    Then, we consider the trajectory of \eqref{eq: system} under the constant control $Bu\equiv\mu b=-\mu Aw$, and conclude by applying Lemma~\ref{lemma: master} with $\mu$ sufficiently large.

    Let us show that $w\in \mathring K$ and estimate its margin.
    Consider $P:=A_0^{-1}A_1$, so that $A=A_0(\idty+P)$ and $\|P\|\le p$.
    By \eqref{eq: theta} and \eqref{eq: A1 condition}, $p<\delta^*/(1+\delta^*)\le1/2$.
    In particular, $\idty+P$ is invertible by the Neumann series \cite{horn2013matrix}, and $w=(\idty+P)^{-1}w_0$.
    Moreover
    \begin{equation} \label{eq: displacement}
        \|w-w_0\|=\bigl\|(\idty+P)^{-1}Pw_0\bigr\|\ \le\ \frac{\|P\|}{1-\|P\|}\|w_0\|\ \le\ \vartheta\|w_0\| ,
    \end{equation}
    so that $(1-\vartheta)\|w_0\|\le\|w\|\le(1+\vartheta)\|w_0\|$.
    Since the left-hand side of \eqref{eq: direction perturbed} is positive, we have $\hat\delta_0>0$, that is, $\vartheta<\delta_0$.
    It follows that $w\in \mathcal B(w_0,\delta_0\|w_0\|)$, which is contained in $K$ by definition of $\delta_0$.
    Hence,
    \begin{equation*}
        \dist\bigl(w,\R^n\setminus K\bigr)\ \ge\ (\delta_0-\vartheta)\|w_0\|\ >\ 0 .
    \end{equation*}
    In particular, $w\in\mathring K$.
    Dividing by $\|w\|\le(1+\vartheta)\|w_0\|$ and using $\delta_0\le1$ gives
    \begin{equation} \label{eq: perturbed equilibrium}
        \delta_K(w)\ \ge\ \hat\delta_0\ \in\ (0,1] .
    \end{equation}

    Consider now the trajectory $x$ of \eqref{eq: system} under the constant control $Bu\equiv\mu b=-\mu Aw$.
    The variation of constants formula \eqref{eq: voc} gives
    \begin{equation} \label{eq: duhamel steering}
        x(t)=\mu w+\varrho(t) ,
        \qquad
        \varrho(t):=e^{At}\bigl(x(0)-\mu w\bigr)+\int_0^te^{A(t-s)}\phi(s)\,ds ,
        \qquad t\ge0 .
    \end{equation}
    By \eqref{eq: perturbed decay}, $\|x(0)\|\le\rho$ and $\|\phi\|\le\bar\phi$, for every $t\ge T$,
    \begin{equation} \label{eq: residual total}
        \|\varrho(t)\|\ \le\ {\mu M_0e^{-\alpha_1T}\|w\|}+{M_0\rho+\frac{M_0\bar\phi}{\alpha_1}} .
    \end{equation}
    Let us now apply Lemma~\ref{lemma: master} at each $t\ge T$, with  $s=s_0=\mu$, $r=\varrho(t)$, $\delta=\hat\delta_0$, $c_1=M_0e^{-\alpha_1T}$ and $c_2=M_0\rho+M_0\bar\phi/\alpha_1$.
    The choices for $\delta$ and $r$ are admissible thanks to \eqref{eq: perturbed equilibrium} and \eqref{eq: residual total}.
    Condition \eqref{eq: direction perturbed} is exactly $c_1<\hat\delta_0$.
    In order to apply Lemma~\ref{lemma: master}, it remains to check \eqref{eq: master condition}, which is implied by $\mu\ge \mu_0$, where
    \begin{equation} \label{eq: mu star perturbed}
        \mu_0 := \max\left\{1,\ \frac{M_0\rho+M_0\bar\phi/\alpha_1}{\bigl(\hat\delta_0-M_0e^{-\alpha_1T}\bigr)(1-\vartheta)\|w_0\|}\right\}.
    \end{equation}
    Indeed, since $\mu\ge\mu_0\ge1$ and $\|w\|\ge(1-\vartheta)\|w_0\|$, we have
    \begin{equation*}
        \mu\bigl(\hat\delta_0-c_1\bigr)\|w\|\ \ge\ \mu\bigl(\hat\delta_0-c_1\bigr)(1-\vartheta)\|w_0\|\ \ge\ c_2 .
    \end{equation*}
    Lemma~\ref{lemma: master} therefore gives the first part of \eqref{eq: steering conclusion}.
    Finally, \eqref{eq: duhamel steering} and \eqref{eq: perturbed decay} give
    \begin{equation*}
        \|x(t)\|\ \le\ \mu\|w\|+M_0\bigl(\rho+\mu\|w\|\bigr)+\frac{M_0\bar\phi}{\alpha_1}\ \le\ R^0_\mu ,
        \qquad 
        \forall t\ge0 ,
    \end{equation*}
    using $\|w\|\le(1+\vartheta)\|w_0\|$ once more.
    Since $R^0_\mu\ge M_0\rho\ge\rho$, the system can be steered into $K$ in time $T$ in the sense of Definition~\ref{def: steering}, with $\rho'=R^0_\mu$.
\end{proof}

We now show that adding some assumptions on the linear part allows one to steer in arbitrary time.
The first case we consider is when $A$ is known and the pair $(A,B)$ is controllable.
Here, given an amplitude $\mu>0$ and a direction $b$ with equilibrium $w=-A^{-1}b$, we steer to $\mu w$ at time $T$ and hold the state there, using
\begin{equation} \label{eq: gramian control}
    u(t)=\begin{cases}
        \mu\,B^\top e^{A^\top(T-t)}\mathcal{G}(T)^{-1}w , & t\in[0,T) , \\
        u_b ,                                             & t\ge T ,
    \end{cases}
\end{equation}
where $u_b\in\R^m$ is any vector with $Bu_b=\mu b$, and $\mathcal{G}(T)$ is the controllability Gramian,
\begin{equation}
    \mathcal{G}(T):=\int_0^{T}e^{As}BB^\top e^{A^\top s}\,ds.
\end{equation}
Since $(A,B)$ is controllable, $\mathcal{G}(T)$ is invertible and \eqref{eq: gramian control} steers the system to $\mu w$ at time $T$ \cite{sontag2013mathematical,trelat2005controle,coron2007control}.
Associated with the Gramian is the constant
\begin{equation}\label{eq: cB}
    c_B(T):=\sup_{0\le t\le T}\bigl\|\mathcal{G}(t)\,e^{A^\top(T-t)}\mathcal{G}(T)^{-1}\bigr\| .
\end{equation}
That is, $c_B(T)$ is the maximal gain along the minimum-energy trajectory steering the origin to a unit vector in time $T$.
It is finite by continuity of $t\mapsto\mathcal{G}(t)e^{A^\top(T-t)}$.

Our result is then the following.

\begin{lemma}[Steering in arbitrary time, controllable case] \label{lemma: controllable steering}
    Let $A_1=0$, so that $A=A_0$ is Hurwitz, assume that $(A,B)$ is controllable, let $K\subset\R^n$ be closed and coradiant with nonempty interior, and assume $\delta^*_K(A)>0$.
    Then \eqref{eq: system} can be steered into $K$ in \emph{every} time $T>0$.
    More precisely, there exists $b\in\Ima(B)$ such that, for every $T>0$ and every $\rho>0$, there exists $\mu_0\ge1$ such that, for every $\mu\ge\mu_0$, the control \eqref{eq: gramian control} drives every solution of \eqref{eq: system} with $\|x(0)\|\le \rho$ so that \eqref{eq: steering conclusion} holds with
    \begin{equation}
        \rho':=\mu\bigl(1+c_B(T)\bigr)\|A^{-1}b\|+(1+M_0)\left(M_0\rho+\frac{M_0\bar\phi}{\alpha_0}\right) .
    \end{equation}
\end{lemma}

\begin{proof}
    Let $b\in\Ima(B)$ be such that $w:=-A^{-1}b\in\mathring K$, which exists since $\delta^*_K(A)>0$, and set $\delta:=\delta_K(w)$.
    The open-loop part of \eqref{eq: gramian control} places the state at $\mu w$ at time $T$, up to the contributions of the initial condition and of $\phi$.
    More precisely, at $t=T$ the variation of constants formula \eqref{eq: voc} yields $\|x(T)-\mu w\|\le M_0\rho+M_0\bar\phi/\alpha_0$.
    For $t\ge T$ the control is the constant $Bu\equiv\mu b$, so, similarly to \eqref{eq: duhamel steering}, we have
    \begin{equation}
        x(t)=\mu w+\varrho(t), \qquad \varrho(t)=e^{A(t-T)}(x(T)-\mu w)+\int_{T}^te^{A(t-s)}\phi(s)\,ds,
        \qquad t\ge T,
    \end{equation}
    hence
    \begin{equation} \label{eq: controllable residual}
        \|\varrho(t)\|\ \le\ (1+M_0)\left(M_0\rho+\frac{M_0\bar\phi}{\alpha_0}\right) , \qquad t\ge T .
    \end{equation}
    This is the form \eqref{eq: master shape} required by Lemma~\ref{lemma: master}, with $s=s_0=\mu$, $r=\varrho(t)$, $\delta=\delta_K(w)$, $c_1=0$ and $c_2$ the right-hand side of \eqref{eq: controllable residual}.
    Condition \eqref{eq: master condition}, together with $s_0\ge1$, then holds for every $\mu\ge\mu_0$, where
    \begin{equation} \label{eq: mu star controllable}
        \mu_0:=\max\left\{1,\ \frac{(1+M_0)\bigl(M_0\rho+M_0\bar\phi/\alpha_0\bigr)}{\delta\,\|w\|}\right\} .
    \end{equation}
    Lemma~\ref{lemma: master}, applied at each $t\ge T$, therefore gives $x(t)\in K$ for all $t\geq T$. 
    As for the state bound, on $[0,T]$ Duhamel's formula and \eqref{eq: gramian control} give $\|x(t)\|\le M_0\rho+M_0\bar\phi/\alpha_0+\mu c_B(T)\|w\|$, where $c_B(T)$ is defined in \eqref{eq: cB}, while on $[T,+\infty)$ we have $\|x(t)\|\le\mu\|w\|+\|\varrho(t)\|$.
    Both are at most $\rho'$.
\end{proof}

When the linear part is not fully known, the minimal time can be removed if the known part is $A_0 = -\lambda\idty$ for some $\lambda > 0$, although the conditions on the unknown part are stronger than in Lemma~\ref{lemma: steering perturbed}.
We observe that in this case, $M_0=1$ and $\alpha_0=\lambda$, so that $p=\bar a/\lambda$ and $\alpha_1=\lambda-\bar a$ in \eqref{eq: theta}.

\begin{lemma}[Steering in arbitrary time, scalar known part] \label{lemma: scalar steering}
    Let $A_0=-\lambda\idty$ with $\lambda>0$.
    Let $K\subset\R^n$ be closed and coradiant with nonempty interior.
    Set $\delta^*:=\delta^*_K(A_0)$ and assume that
    \begin{equation} \label{eq: A1 scalar condition}
        \bar a\ <\ \lambda\cdot\frac{\delta^*}{2+\delta^*} .
    \end{equation}
    Then \eqref{eq: system} can be steered into $K$ in \emph{every} time $T>0$.
    More precisely, there exists $b\in\Ima(B)$ such that, for every $T>0$ and every $\rho>0$, there exists $\mu_0\ge1$ such that, for every $\mu\ge\mu_0$, the constant control $Bu\equiv\mu b$ drives every solution of \eqref{eq: system} with $\|x(0)\|\le\rho$ so that \eqref{eq: steering conclusion} holds with $\rho'=R^0_\mu$ defined in \eqref{eq: state bound perturbed}.
\end{lemma}

\begin{proof}
    Let $b\in\Ima(B)$ be such that $w_0:=b/\lambda\in\mathring K$ and $\delta_0:=\delta_K(w_0)>2\vartheta$.
    Such a $b$ exists by definition of the best margin, since \eqref{eq: A1 scalar condition} is equivalent to $2\vartheta<\delta^*$.
    By \eqref{eq: A1 scalar condition} we also have $\bar a<\lambda\delta^*/(1+\delta^*)$, which is the hypothesis \eqref{eq: A1 condition} of Lemma~\ref{lemma: steering perturbed}.
    Since moreover $\vartheta<\delta_0/2<\delta_0$, the same argument as in the proof of the latter places $w:=-A^{-1}b$ in $\mathring K$, with
    \begin{equation} \label{eq: scalar equilibrium}
        \delta_K(w)\ \ge\ \hat\delta_0=\frac{\delta_0-\vartheta}{1+\vartheta} ,
        \qquad
        \|w\|\ \ge\ (1-\vartheta)\frac{\|b\|}{\lambda} .
    \end{equation}
    Moreover $\vartheta<1/2$ gives $p<1/3$, hence $\alpha_1=\lambda(1-p)>2\lambda/3$.

    Since $A_0=-\lambda\idty$ commutes with $A_1$, we have $e^{At}=e^{-\lambda t}e^{A_1t}$, so that $\|e^{At}\|\le e^{-(\lambda-\|A_1\|)t}\le e^{-\alpha_1t}$.
    Under the constant control $Bu\equiv\mu b$, the variation of constants formula \eqref{eq: voc} and $\int_0^te^{A(t-s)}ds=A^{-1}(e^{At}-\idty)$ give, for any $t\ge 0$,
    \begin{equation} \label{eq: duhamel scalar}
        x(t)=\mu\bigl(\idty-e^{At}\bigr)w+\varrho(t),
        \quad
        \varrho(t):=e^{At}x(0)+\int_0^te^{A(t-s)}\phi(s)\,ds.
    \end{equation}
    In particular, the residual $\varrho$ is bounded uniformly in $t$ and in $\mu$ by $\|\varrho(t)\|\le\rho+{\bar\phi}/{\alpha_1}$.

    It therefore suffices to keep the curve $t\mapsto(\idty-e^{At})w$ far enough inside $K$ for every $t\ge T$.
    To this end, we split the curve into a part along the ray $\R_+w$ and a remainder $z(t)$, as follows:
    \begin{equation} \label{eq: split}
        \bigl(\idty-e^{At}\bigr)w=\bigl(1-e^{-\lambda t}\bigr)w+z(t),
        \qquad
        z(t):=e^{-\lambda t}\bigl(\idty-e^{A_1t}\bigr)w .
    \end{equation}
    From $\|\idty-e^{A_1t}\|\le e^{\|A_1\|t}-1\le e^{\bar at}-1$  and $\bar a<\lambda$, we can bound the remainder for $t\ge0$ as
    \begin{equation} \label{eq: z bound}
        \begin{split}
            \|z(t)\|\ &\le\ e^{-\lambda t}\bigl(e^{\bar at}-1\bigr)\|w\|\ =\ \bar a\|w\|\int_0^te^{\bar a\tau-\lambda t}\,d\tau \\
            &\le\ \bar a\|w\|\int_0^te^{-\lambda(t-\tau)}\,d\tau\ =\ \bigl(1-e^{-\lambda t}\bigr)\frac{\bar a}{\lambda}\,\|w\| .
        \end{split}
    \end{equation}
    Both terms of \eqref{eq: split} carry the same factor $1-e^{-\lambda t}$, so their comparison is independent of $t$ and of $\mu$, and no minimal time is needed.

    In order to conclude the proof, we now apply Lemma~\ref{lemma: master} to the trajectory \eqref{eq: duhamel scalar}.
    Let us check that the hypotheses are satisfied.
    Substituting \eqref{eq: split} into \eqref{eq: duhamel scalar} gives $x(t)=\mu\bigl(1-e^{-\lambda t}\bigr)w+\bigl(\mu z(t)+\varrho(t)\bigr)$, which by \eqref{eq: z bound} has the form \eqref{eq: master shape} required by Lemma~\ref{lemma: master}, with  $s=\mu(1-e^{-\lambda t})\ge s_0:=\mu(1-e^{-\lambda T})$, $r=\mu z(t)+\varrho(t)$, $c_1=\bar a/\lambda=\vartheta/(1+\vartheta)$ and $c_2=\rho+\bar\phi/\alpha_1<\rho+3\bar\phi/(2\lambda)$
    for $t\ge T$, since $t\mapsto1-e^{-\lambda t}$ is increasing.
    Moreover, by \eqref{eq: scalar equilibrium} we can take $\delta=\hat\delta_0$.
    Recalling that $2\vartheta<\delta_0$, we have
    \begin{equation}
        \delta-c_1\ \ge\ \frac{\delta_0-\vartheta}{1+\vartheta}-\frac{\vartheta}{1+\vartheta}\ =\ \frac{\delta_0-2\vartheta}{1+\vartheta}\ >\ 0.
    \end{equation}
    With $\|w\|\ge(1-\vartheta)\|b\|/\lambda$, the condition \eqref{eq: master condition} is implied by
    \begin{equation*}
        \mu\bigl(1-e^{-\lambda T}\bigr)\frac{(\delta_0-2\vartheta)(1-\vartheta)}{1+\vartheta}\cdot\frac{\|b\|}{\lambda}\ \ge\ \rho+\frac{3\bar\phi}{2\lambda} ,
    \end{equation*}
    and the requirement $s_0\ge1$ is $\mu(1-e^{-\lambda T})\ge1$.
    Both hold for every $\mu\ge\mu_0$, where
    \begin{equation} \label{eq: mu star scalar}
        \mu_0:=\frac{1}{1-e^{-\lambda T}}\max\left\{1,\ \
        \frac{\lambda(1+\vartheta)\bigl(\rho+3\bar\phi/(2\lambda)\bigr)}{(1-\vartheta)(\delta_0-2\vartheta)\,\|b\|}\right\} .
    \end{equation}

    Finally, Lemma~\ref{lemma: master} applied at each $t\ge T$ with $s=\mu\bigl(1-e^{-\lambda t}\bigr)$ gives  $x(t)\in K$ for all $t\ge T$, and $\|x(t)\|\le R^0_\mu$ for every $t\ge0$ follows from $\|e^{At}\|\le1$ and $M_0=1$.
\end{proof}

\subsection{Concatenating the steps} \label{subsec: concatenation}

In this section we show that the previous results can be concatenated to sequentially steer the system into a sequence of coradiant sets.

\begin{lemma}
    \label{lemma: sequential}
    Let $S$ be a  coradiant saturation. 
    For each $i\in\llb1,n\rrb$, let \eqref{eq: system} be steerable into $K_i$ in some time $T_i>0$ in the sense of Definition~\ref{def: steering}.
    Then for every $\rho>0$, and $T,\tau>0$ such that $T\ge n\tau+\sum_{i=1}^nT_i$, there exist intervals $J_i$ with $|J_i|=\tau$ such that $\mathcal{E}=\{(v_i,J_i)\}_{i\in\llb1,n\rrb}$ is an excitation design uniformly realizable in time $T$ and radius $\rho$.
    Moreover, there is $\bar\rho\ge\rho$ such that every solution of \eqref{eq: system} with $\|x(0)\|\le\rho$ under the control corresponding to $\mathcal E$ satisfies $\|x(t)\|\le\bar\rho$ for every $t\in[0,T]$.
\end{lemma}

\begin{proof}
    Set $s_0:=0$ and define the times
    \begin{equation} \label{eq: window spacing}
        t_1:=T_1, \qquad t_i:=s_{i-1}+T_i , \qquad s_i:=t_i+\tau , \quad i\in\llb1,n\rrb ,
    \end{equation}
    and $J_i:=[t_i,s_i]$, so that $t_1\ge T_1$, $t_i\ge s_{i-1}+T_i$ and $s_n=n\tau+\sum_{i=1}^nT_i\le T$.
    These intervals are pairwise disjoint, and the $v_i$ are linearly independent since $S$ is a coradiant saturation, so that \eqref{eq: rank condition} holds and $\mathcal{E}:=\{(v_i,J_i)\}_{i\in\llb1,n\rrb}$ is an excitation design.
    The construction of the control is by induction on $i$, the induction hypothesis being a bound $\rho_i$ on the norm of the state at time $s_{i-1}$, with $\rho_1:=\rho$.
    Given $\rho_i$, Definition~\ref{def: steering} applied with the initial time $s_{i-1}$ furnishes for $K_i$ and the radius $\rho_i$ a control $u_i$ and a radius $\rho'_i$ such that any solution issued at time $s_{i-1}$ from a state of norm at most $\rho_i$ satisfies $x(t)\in K_i$ for every $t\ge s_{i-1}+T_i$, hence for every $t\in J_i$ by \eqref{eq: window spacing}, and $\|x(t)\|\le\rho'_i$ throughout.
    By \eqref{eq: saturation on cones}, $S(x(t))=v_i$ on $J_i$, which is the $i$-th requirement for the excitation design, see \eqref{eq: star}.
    Setting $\rho_{i+1}:=\rho'_i$ closes the induction, and the concatenation $u:=\sum_{i=1}^nu_i\mathbf{1}_{[s_{i-1},s_i)}$, extended to $[0,T]$ by $u_n$ on $[s_n,T]$, is the desired control.
    Finally, $\rho\le\rho'_1\le\cdots\le\rho'_n$ by Definition~\ref{def: steering}, so $\bar\rho:=\rho'_n$ bounds $\|x(t)\|$ on $[0,T]$.
\end{proof}

\subsection{Proofs of the theorems} \label{subsec: proofs theorems}

We are finally in a position to prove the main results of Section~\ref{subsec: structural}.

\begin{proof}[Proof of Theorem~\ref{thm:perturbed-A}]
    For each $i$ let $\delta_i:=\delta^*_{K_i}(A_0)>0$, and set
    \begin{equation} \label{eq: epsilon}
        \varepsilon_0:=\frac{\alpha_0}{M_0}\min_{1\le i\le n}\frac{\delta_i}{1+\delta_i} .
    \end{equation}
    Let $\bar a<\varepsilon_0$.
    Then $\bar a$ satisfies \eqref{eq: A1 condition} with $\delta^*=\delta_i$ for every $i$.
    Lemma~\ref{lemma: steering perturbed} therefore applies to each $K_i$, and makes \eqref{eq: system} steerable into $K_i$ in every time larger than
    \begin{equation*}
        T_{\min}^i:=\frac{1}{\alpha_1}\ln\frac{M_0}{\hat\delta_i} ,
        \qquad
        \hat\delta_i:=\frac{\delta_i-\vartheta}{1+\vartheta}\ >\ 0 ,
    \end{equation*}
    with $\vartheta$ and $\alpha_1=\alpha_0-M_0\bar a$ as in \eqref{eq: theta}.
    Set
    \begin{equation} \label{eq: T star}
        T^*:=\sum_{i=1}^nT_{\min}^i=\sum_{i=1}^n\frac{1}{\alpha_0-M_0\bar a}\ln\frac{M_0}{\hat\delta_i} .
    \end{equation}
    Given $T>T^*$, we can choose times $T_i>T_{\min}^i$ with $\sum_iT_i<T$, and Lemma~\ref{lemma: sequential} with $\tau:=(T-\sum_iT_i)/n>0$ concludes.
    By Lemma~\ref{lemma: steering perturbed}, the directions, the times $T_i$ and the amplitudes are all computed from $A_0$, $B$, the $K_i$, $\bar\phi$, $\rho$, $T$ and $\bar a$.
    So are the intervals $J_i$ and the radii $\rho_i$ of the proof of Lemma~\ref{lemma: sequential}, hence the design $\mathcal{E}$, the radius $\bar\rho$ and the control $u$, which is piecewise constant since each step uses a constant control.
    In particular, neither depends on $A_1$ or $\phi$ within the bounds $\|A_1\|\le\bar a$ and $\|\phi\|_\infty\le\bar\phi$.
\end{proof}

The time \eqref{eq: T star} increases with the bound $\bar a$ and blows up as $\bar a\to\varepsilon_0$.
For $\bar a=0$, it reduces to $\sum_{i=1}^n\alpha_0^{-1}\ln(M_0/\delta_i)$.

\begin{proof}[Proof of Theorem~\ref{thm: arbitrary time}, case \emph{i.}]
    Here $\bar a=0$, so $A_1=0$ and $A=A_0$ is known and Hurwitz and the hypothesis of Theorem~\ref{thm:perturbed-A} reads $\Ima(A^{-1}B)\cap\mathring K_i\neq\varnothing$ for each $i$.
    Since $(A,B)$ is controllable, Lemma~\ref{lemma: controllable steering} makes \eqref{eq: system} steerable into $K_i$ in \emph{every} time $T_i>0$.
    Given $T>0$, taking the $T_i$ small enough that $\sum_iT_i<T$ and applying Lemma~\ref{lemma: sequential} with $\tau:=(T-\sum_iT_i)/n>0$ concludes.
    Hence $T^*=0$, and by Lemma~\ref{lemma: controllable steering} the control on the $i$-th step is computed from $A_0$, $B$, $K_i$, $\bar\phi$, $\rho$ and $T_i$.
\end{proof}

\begin{proof}[Proof of Theorem~\ref{thm: arbitrary time}, case \emph{ii.}]
    Here $A_0=-\lambda\idty$ with $\lambda>0$, so $M_0=1$, $\alpha_0=\lambda$ and $A=-\lambda\idty+A_1$ has the structure required by Lemma~\ref{lemma: scalar steering}.
    With $\delta_i:=\delta^*_{K_i}(A_0)>0$ as in the proof of Theorem~\ref{thm:perturbed-A}, set
    \begin{equation} \label{eq: epsilon scalar}
        \varepsilon_0:=\lambda\min_{1\le i\le n}\frac{\delta_i}{2+\delta_i} ,
    \end{equation}
    which depends only on $\lambda$, $B$ and the $K_i$.
    Since $\bar a<\varepsilon_0$, we have $\bar a<\lambda\delta_i/(2+\delta_i)$ for every $i$, which is \eqref{eq: A1 scalar condition} for $K_i$.
    Lemma~\ref{lemma: scalar steering} then makes \eqref{eq: system} steerable into $K_i$ in every time $T_i>0$, and one concludes as in case \emph{i.}
    Hence $T^*=0$, and by Lemma~\ref{lemma: scalar steering} the control on the $i$-th step is computed from $\lambda$, $B$, $K_i$, $\bar\phi$, $\rho$ and $\bar a$.
\end{proof}

We conclude this section with the proof of Proposition~\ref{prop:rank-B}.

\begin{proof}[Proof of Proposition~\ref{prop:rank-B}]
    We first show openness.
    Let \eqref{eq: condition A_0} hold for $A_0$ with sets $K_1,\ldots,K_n$, and pick $b_k\in\Ima B$ with $A_0^{-1}b_k\in\mathring K_k$.
    Since $A'_0\mapsto A_0'^{-1}b_k$ is continuous near $A_0$ and $\mathring K_k$ is open, $A_0'^{-1}b_k\in\mathring K_k$ for every $A'_0$ close enough to $A_0$, which can moreover be taken Hurwitz since being Hurwitz is an open condition.

    We now show density, that is, that any Hurwitz $A_0$ has a Hurwitz $A'_0$ arbitrarily close to it for which \eqref{eq: condition A_0} holds.
    Let $b,b'\in\Ima B$ be linearly independent, and set $Z_0:=A_0^{-1}[b\ b']\in\R^{n\times2}$, which has rank $2$.
    For $Z\in\R^{n\times2}$ with rows $\omega_1^\top,\ldots,\omega_n^\top$, let $p(Z):=\prod_{i<j}\det[\omega_i\ \omega_j]$.
    Then, $p$ is a polynomial in the entries of $Z$ which is non-zero.
    Indeed, $p(\bar Z)\neq 0$ for $\bar Z$ with rows $\bar\omega_i=(1,i)$.
    Since the zero set of a non-zero polynomial has empty interior, there exists $Z$ arbitrarily close to $Z_0$ such that $p(Z)\neq0$.
    In particular, the rows of $Z$ are pairwise linearly independent, and thus $Z^\top Z$ is invertible.
    Fix any such $Z$ and consider $A'_0:=A_0+A_0(Z_0-Z)Z^\dagger$, with $Z^\dagger=(Z^\top Z)^{-1}Z^\top$.
    We have $A'_0Z=[b\ b']$ and $\|A'_0-A_0\|\le\|A_0\|\,\|Z_0-Z\|\,\|Z^\dagger\|$.
    As $Z^\dagger$ stays bounded near $Z_0$ and being Hurwitz is an open condition, it follows that $A'_0$ is Hurwitz and arbitrarily close to $A_0$.
    In particular, $A'_0$ is invertible and $\Ima Z\subset\Ima(A_0'^{-1}B)$.

    Consider now the points $Zy$, whose coordinates are $x_i=\langle\omega_i,y\rangle$, for $y$ on the unit circle of $\R^2$.
    Each $x_i$ vanishes at exactly two antipodal points, and by the choice of $Z$ these $2n$ points are distinct.
    They split the circle into $2n$ open arcs, on each of which no coordinate vanishes, so that $Zy$ stays in the interior of a single orthant $\mathcal O_{\mathbf s}$.
    Crossing the zero of $x_i$ changes the sign of $x_i$ only, so that the patterns of two adjacent arcs differ by $\pm e_i$.
    Since every $i\in\llb1,n\rrb$ occurs, the span of these $2n$ patterns contains $e_1,\ldots,e_n$, and we can extract $n$ linearly independent patterns $\mathbf s_1,\ldots,\mathbf s_n$.
    Finally, if $y$ lies on an arc with pattern $\mathbf s_k$, then $\lambda Zy\in\mathring K_{\mathbf s_k}$ for $\lambda$ large enough.
    Hence \eqref{eq: condition A_0} holds for $A'_0$ with $K_k:=K_{\mathbf s_k}$.%
\end{proof}

\section{Output feedback tracking} \label{sec: feedback proofs}

In this section we prove the statements of Section~\ref{subsec: state}, namely Theorem~\ref{thm: output feedback}.

\begin{figure}%
    \centering
    \begin{tikzpicture}[scale=0.85,>=Stealth]
        \def\wo{0.9}
        \def\hi{1.5}
        \def\xL{-1.2}
        \def\xR{6.0}

        \fill[gray!10] (\wo,-\hi-0.35) rectangle (5.0-\wo,\hi+0.35);
        \node[gray!65,font=\scriptsize,anchor=south] at (2.5,\hi+0.30) {proportional range};

        \draw[->,black!70] (\xL,0) -- (\xR,0) node[right,font=\small] {$s$};
        \draw[->,black!70] (0,-\hi-0.55) -- (0,\hi+0.75) node[above,font=\small] {$\eta_0$};

        \draw[densely dotted,gray!60] (0,\hi) -- (\xL,\hi);
        \draw[densely dotted,gray!60] (5.0,-\hi) -- (\xR,-\hi);
        \node[font=\scriptsize,anchor=east] at (\xL+0.02,\hi) {$1$};
        \node[font=\scriptsize,anchor=west] at (\xR-0.02,-\hi) {$-1$};

        \draw[densely dashed, gray!55] (\wo,0) -- (\wo,\hi*0.62);
        \draw[densely dashed,gray!55] (5.0-\wo,0) -- (5.0-\wo,-\hi*0.62);
        \draw[densely dashed,gray!55] (5.0,0) -- (5.0,-\hi);

        \node[font=\scriptsize,anchor=north] at (\wo,-0.06) {$\omega$};
        \node[font=\scriptsize,anchor=south] at (5.0-\wo,0.06) {$1-\omega$};
        \node[font=\scriptsize,anchor=south west] at (5.0,0.06) {$1$};

        \draw[very thick,blue!65!black] (\xL,\hi) -- (0,\hi) -- (\wo,\hi*0.62)
        .. controls (\wo+0.25,0.4) and (2.2,0.12) .. (2.72,0)
        .. controls (3.24,-0.12) and (5.0-\wo-0.2,-0.45) .. (5.0-\wo,-\hi*0.62)
        -- (5.0,-\hi) -- (\xR,-\hi);

        \fill[black] (2.72,0) circle (1.3pt);
        \node[font=\scriptsize,anchor=south,inner sep=2pt] at (2.72,0.04) {$S_0(\gamma)$};

        \node[font=\scriptsize,blue!65!black,anchor=south west,inner sep=1pt] at (0.42,\hi*0.81+0.04) {$\ell_1$};
        \node[font=\scriptsize,blue!65!black,anchor=north east,inner sep=1pt] at (5.0-0.42,-\hi*0.81-0.04) {$\ell_2$};
    \end{tikzpicture}
    \caption{The scalar feedback \eqref{eq: eta feedback} as a function of the measured coordinate $s$.
    }
    \label{fig: feedback}
\end{figure}
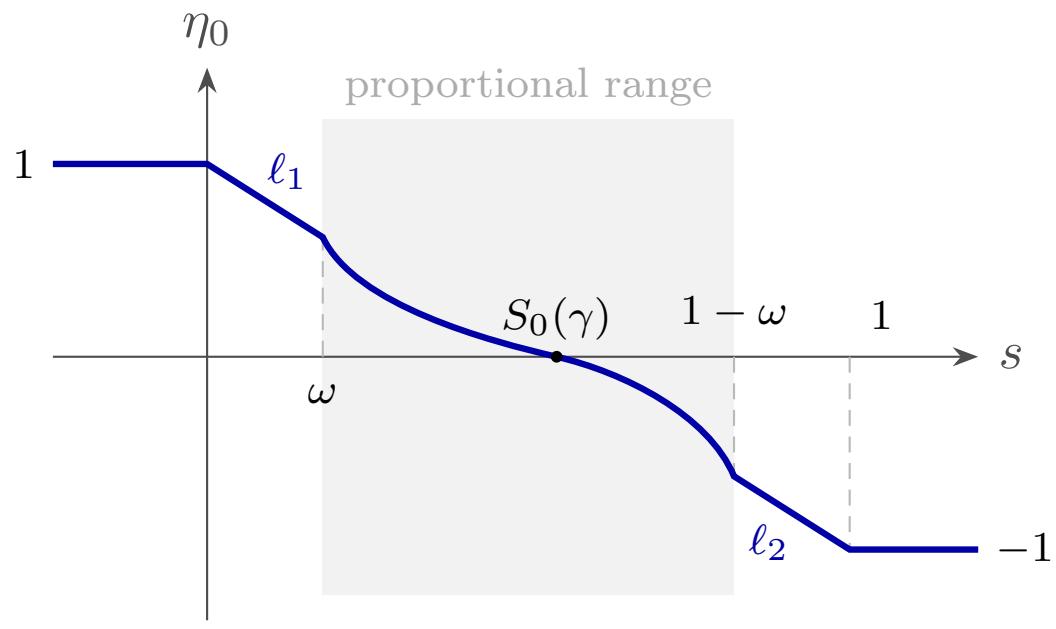

Observe that reading the output coordinate by coordinate tells whether each coordinate of the state is saturated.
This suggests the following coordinate-wise strategy for the output feedback loop:
In a saturated coordinate the law pushes the state back with a constant effort, while in an unsaturated one it switches to proportional tracking.
The scalar feedback is obtained by scalar multiplication of the following function, where $\omega\in(0,1/2)$ and $\gamma\in\R$:
\begin{equation} \label{eq: eta feedback}
    \eta_0(s;\gamma,\omega)=
    \begin{cases}
        1,                                     & s\leq0,                  \\
        \ell_1(s),               & 0<s<\omega,              \\
        -\bigl(S_0^{-1}(s)-\gamma\bigr), & \omega\leq s\leq1-\omega, \\
        \ell_2(s),               & 1-\omega<s<1,            \\
        -1,                                    & s\geq1,
    \end{cases}
\end{equation}
where $\ell_1$ and $\ell_2$ are affine functions making $\eta_0$ continuous.
Figure~\ref{fig: feedback} shows the five branches.
The actual output feedback control is then defined by a scalar gain $k>0$ and a reference value $\gamma\in\R^n$ as
\begin{equation} \label{eq: feedback law}
    \eta_k(y,\gamma):=k\boldsymbol\eta_0\bigl(Hy;\gamma,\omega\bigr) ,
    \quad 
   \boldsymbol\eta_0(s;\gamma,\omega):=\bigl(\eta_0(s_1;\gamma_1,\omega),\ldots,\eta_0(s_n;\gamma_n,\omega)\bigr)^\top.
\end{equation}

For $\omega\in(0,1/2)$ write
\begin{equation}
    \Omega_{\omega}:=\bigl[S_0^{-1}(\omega),\,S_0^{-1}(1-\omega)\bigr]^{n}\subset\mathring\Omega
\end{equation}
for the box obtained from $\Omega$ by staying at distance $\omega$ from the saturation levels, measured on the output side, and, for $L>0$ and $X\subset\R^n$, let $C^{1,L}(\R,X)$ denote the set of $C^1$ curves with values in $X$ whose derivative is bounded by $L$.

\begin{lemma}
    \label{lemma: tracking}
    Let $g:\R^n\times\R^m\times\R\to\R^n$ be continuous with
    \begin{equation} \label{eq: linear growth}
        \|g(x,u,t)\|\le a\|x\|+c_g , \qquad (x,u,t)\in\R^n\times\R^m\times\R ,
    \end{equation}
    for some \emph{known} constants $a,c_g>0$, and consider
    \begin{equation} \label{eq: feedback system}
        \dot x=g(x,u,t)+u, \qquad y=WS(x),
    \end{equation}
    with $W$ known and injective, $S$ a componentwise saturation and $H$ as in \eqref{eq: componentwise output}.
    Let $\varepsilon,\rho,L>0$, let $0<T_1<T_2$, let $\omega\in(0,1/4)$ and let $\gamma\in C^{1,L}(\R,\Omega_{2\omega})$.
    Let $\nu:[0,+\infty)\to\R^q$ be measurable with $\|H\nu(t)\|_\infty\le\delta_H\le\omega/2$ for all $t$, where $\delta_H=0$ unless $S_0^{-1}$ is locally Lipschitz on $(0,1)$.
    Then there exists $k^\star>0$ such that for every gain $k>k^\star$ the solutions of \eqref{eq: feedback system} under the control $u(t)=\eta_k(y(t)+\nu(t),\gamma(t))$ of \eqref{eq: feedback law}, with $\|x(0)\|\le\rho$, are global in time and satisfy
    \begin{equation} \label{eq: tracking conclusion}
        x(t)\in\Omega_{\omega} \quad \forall t\geq T_1,
        \qquad
        \|x(t)-\gamma(t)\|\leq\varepsilon+\sqrt n\,\ell_\omega\delta_H \quad \forall t\geq T_2 ,
    \end{equation}
    where $\ell_\omega$ depends only on $S$ and  $\omega$.
\end{lemma}

\begin{proof}
    Continuity of $S$, $g$ and $\eta_0$ and measurability of $\nu$ give local existence and absolute continuity of the solutions, and by \eqref{eq: componentwise output} the closed loop reads
    \begin{equation} \label{eq: scalar loop}
        \dot x_j=g_j(x,u,t)+k \ \eta_0\bigl(S_0(x_j)+\tilde\nu_j(t);\gamma_j,\omega\bigr) , \qquad j\in\llb1,n\rrb , \quad \tilde\nu:=H\nu .
    \end{equation}
    Set
    \begin{equation}
       \label{eq: c omega} 
        c_\omega:=\min\Bigl\{1,\ S_0^{-1}(1-\tfrac32\omega)-S_0^{-1}(1-2\omega),\ S_0^{-1}(2\omega)-S_0^{-1}(\tfrac32\omega)\Bigr\}\ >\ 0 .
    \end{equation}
    Observe that the positivity of $c_\omega$ follows from $\omega>0$ and the strict monotonicity of $S_0$ on $[\ubar R,\bar R]$, while $\omega<1/4$ makes $\Omega_{2\omega}$ nonempty.
    For every $t$ and $|r|\le\omega/2$, we claim that
    \begin{equation} \label{eq: eta sign}
        \eta_0\bigl(S_0(\xi)+r;\gamma_j(t),\omega\bigr)
        \begin{cases}
            \le-c_\omega ,    & \xi\ge S_0^{-1}(1-\omega) , \\
            \ge c_\omega , & \xi\le S_0^{-1}(\omega) .
        \end{cases}
    \end{equation}
    Indeed, recall that $\gamma_j(t)\in[S_0^{-1}(2\omega),S_0^{-1}(1-2\omega)]$, and let $\xi\ge S_0^{-1}(1-\omega)$, so that $s:=S_0(\xi)+r\ge1-\frac32\omega>\omega$.
    If $s\le1-\omega$, then $\eta_0(s)=-(S_0^{-1}(s)-\gamma_j)\le-c_\omega$ by \eqref{eq: c omega}.
    Otherwise, $\eta_0$ is affine on $[1-\omega,1]$ and constant beyond, with $\eta_0(1-\omega)\le-c_\omega$ and $\eta_0(1)=-1\le-c_\omega$.
    The lower level is symmetric.

    Consider now the following quantities
    \begin{gather} \label{eq: Lambda}
        \Lambda:=\max\bigl\{\rho,|\ubar R|,|\bar R|\bigr\} ,
        \qquad
        \kappa:=a\sqrt n\,\Lambda+c_g+L ,\\
        \label{eq: k star}
        k^\star:=\max\left\{
        \frac{1}{c_\omega}\left(\frac{2\Lambda}{T_1}+a\sqrt n\,(\Lambda+1)+c_g\right) ,\ \
        \frac{2\sqrt n\,\kappa}{\varepsilon} ,\ \
        \frac{1}{T_2-T_1}\ln\frac{2\sqrt n\,(\bar R-\ubar R)}{\varepsilon}
        \right\} .
    \end{gather}
    Throughout, $k>k^\star$, whose first term in particular gives
    \begin{equation} \label{eq: k zero}
        k\ >\ \frac{a\sqrt n\,(\Lambda+1)+c_g}{c_\omega} .
    \end{equation}

    We now deduce from \eqref{eq: eta sign} that $\{\|x\|_\infty\le\Lambda\}$ is positively invariant.
    Fix $c\in(0,1]$, and let $s_1$ be the infimum of the times when $\|x(t)\|_\infty\ge\Lambda+c$. Assuming that $s_1$ is finite, there exists $j$ such that $|x_j(s_1)|=\Lambda+c$, say $x_j(s_1)=\Lambda+c$, the other case being symmetric.
    Since $x_j(0)\le\rho\le\Lambda$, the set of times before $s_1$ with $x_j=\Lambda$ is nonempty and closed.
    Let $s_0$ be its maximum, so that $x_j>\Lambda$ on $(s_0,s_1]$.
    By \eqref{eq: linear growth} and the definition of $\Lambda$, for all $t\in [s_0,s_1]$,
    \begin{equation}
        |g_j(x(t),u(t),t)|\le a\sqrt n(\Lambda+c)+c_g
        \qquad\text{and}\qquad
        x_j(t)\ge\Lambda\ge\bar R\ge S_0^{-1}(1-\omega).
    \end{equation}
    Since $c\le1$, using \eqref{eq: scalar loop}, \eqref{eq: k zero}, and \eqref{eq: eta sign} this gives $\dot x_j(t)\ \le\ a\sqrt n\,(\Lambda+c)+c_g-kc_\omega< 0$ almost everywhere on $[s_0,s_1]$, contradicting $x_j(s_1)>x_j(s_0)$.
    Hence $\|x(t)\|_\infty<\Lambda+c$ as long as the solution is defined, so that the solution is bounded, hence global.
    Since $c\in(0,1]$ is arbitrary, we obtain $\|x(t)\|_\infty\le\Lambda$ for every $t\ge0$, and therefore
    \begin{equation} \label{eq: a priori bound}
        \|g(x(t),u(t),t)\|\ \le\ a\sqrt n\,\Lambda+c_g , \qquad \forall t\ge0 .
    \end{equation}

    We now show that the set $\Omega_\omega$ is reached before $T_1$ and is invariant, which settles the first part of \eqref{eq: tracking conclusion}.
    By \eqref{eq: eta sign} and \eqref{eq: a priori bound}, whenever $x_j(t)\ge S_0^{-1}(1-\omega)$ one has $\dot x_j\le-(kc_\omega-a\sqrt n\Lambda-c_g)<0$, and symmetrically $\dot x_j>0$ whenever $x_j(t)\le S_0^{-1}(\omega)$.
    The interval $[S_0^{-1}(\omega),S_0^{-1}(1-\omega)]$ is therefore positively invariant for $x_j$, and since $|x_j(0)|\le\Lambda$ and $|S_0^{-1}(s)|\le\Lambda$ for $s\in[0,1]$, it is reached after a time at most $\frac{2\Lambda}{kc_\omega-a\sqrt n\,\Lambda-c_g}$, which tends to $0$ as $k\to\infty$ and is at most $T_1$ by the first term of \eqref{eq: k star}.
    Hence $x(t)\in\Omega_{\omega}$ for every $t\ge T_1$, the first assertion of \eqref{eq: tracking conclusion}.

    Let now $t\ge T_1$.
    We have $S_0(x_j(t))\in[\omega,1-\omega]$, so $S_0^{-1}(S_0(x_j))=x_j$ and \eqref{eq: scalar loop} reads $\dot x_j=g_j-k(x_j-\gamma_j)+r_j$,  with $r_j:=k\bigl(\eta_0(S_0(x_j)+\tilde\nu_j)-\eta_0(S_0(x_j))\bigr)$.
    If $S_0^{-1}$ is not locally Lipschitz on $(0,1)$, then $\delta_H=0$, so that $r_j=0$, and we set $\ell_\omega:=0$.
    Otherwise, let $\lambda_\omega$ be a Lipschitz constant of $S_0^{-1}$ on $I_\omega:=[\omega/2,1-\omega/2]$.
    Then $\eta_0$ is $\lambda_\omega$-Lipschitz on $[\omega,1-\omega]$, and affine on $[\omega/2,\omega]$ and $[1-\omega,1-\omega/2]$ with slope at most $(1+\bar R-\ubar R)/\omega$ in absolute value, so that $|r_j|\le k\ell_\omega\delta_H$ with $\ell_\omega:=\max\bigl\{\lambda_\omega,\ (1+\bar R-\ubar R)/\omega\bigr\}$.
    With $e_j:=x_j-\gamma_j$ and $\|\dot\gamma\|\le L$, \eqref{eq: a priori bound} and \eqref{eq: Lambda} give
    \begin{equation}
        D^+|e_j|\ \le\ \kappa+k\ell_\omega\delta_H-k|e_j| \qquad \text{almost everywhere on }[T_1,+\infty) ,
    \end{equation}
    where $D^+$ is the upper right Dini derivative, $|e_j|$ being absolutely continuous but not differentiable at its zeros.
    Hence, by the comparison lemma \cite{khalil1996nonlinear} and $|e_j(T_1)|\le\bar R-\ubar R$,
    \begin{equation}
        |e_j(t)|\ \le\ \bigl(\bar R-\ubar R\bigr)e^{-k(t-T_1)}+\frac{\kappa}{k}+\ell_\omega\delta_H , \qquad t\ge T_1 .
    \end{equation}
    The constant $\kappa$ does not depend on $k$, so the first two terms are at most $\varepsilon/(2\sqrt n)$ for every $t\ge T_2$ as soon as
    \begin{equation}
        k\ >\ \max\left\{\frac{2\sqrt n\,\kappa}{\varepsilon},\ \frac{1}{T_2-T_1}\ln\frac{2\sqrt n\,(\bar R-\ubar R)}{\varepsilon}\right\} ,
    \end{equation}
    whose right-hand side collects the two remaining terms of \eqref{eq: k star}.
    It follows that $\|x(t)-\gamma(t)\|\le\sqrt n\max_j|e_j(t)|\le\varepsilon+\sqrt n\,\ell_\omega\delta_H$ for $t\ge T_2$, concluding the proof.
\end{proof}

We can now turn to the proof of Theorem~\ref{thm: output feedback}.

\begin{proof}[Proof of Theorem~\ref{thm: output feedback}]
    It suffices to prove the second part, since for $\nu=0$ the argument below does not use the Lipschitz continuity of $S_0^{-1}$ and gives the first part.
    Let $\mathcal{K}\subset\mathring\Omega$ be the compact set of the statement and let $L>0$ bound $\|\dot\gamma\|$.
    Since $\mathcal{K}$ is compact and contained in the open box $\mathring\Omega$, there is $\zeta>0$ with $\mathcal{K}\subset[\ubar R+\zeta,\bar R-\zeta]^n$, and then, $S_0^{-1}$ being continuous on $[0,1]$ with $S_0^{-1}(0)=\ubar R$ and $S_0^{-1}(1)=\bar R$, there is some $\omega\in(0,1/4)$ with $\mathcal{K}\subset\Omega_{2\omega}$, so that $\gamma\in C^{1,L}(\R,\Omega_{2\omega})$.
    The system \eqref{eq: system-B-invertible} is of the form \eqref{eq: feedback system} with $g(x,u,t)=Ax+\phi(x,u,t)$, which is continuous and satisfies \eqref{eq: linear growth} with $a=\|A\|>0$ and $c_g=\bar\phi>0$, uniformly in $u$ and $t$.
    We take for $\eta_k$ the feedback \eqref{eq: feedback law} with this $\omega$ and with $H$ as in \eqref{eq: componentwise output}. It depends only on $k$, $W$, $S$, $S_0$ and $\mathcal{K}$.
    Let $\|H\|_\infty$ be the operator norm of $H$ for $\|\cdot\|_\infty$, and set $\delta_0:=\omega/(2\|H\|_\infty)$, so that $\|H\nu(t)\|_\infty\le\delta_H:=\|H\|_\infty\delta\le\omega/2$.
    Lemma~\ref{lemma: tracking} then gives \eqref{eq: tracking conclusion}, and $\Omega_{\omega}\subset\mathring\Omega$ turns it into \eqref{eq: feedback conclusion} with $c_\nu:=\sqrt n\,\ell_\omega\|H\|_\infty$.
    Finally, let $t\ge T_1$, so that $S_0(x_j)\in[\omega,1-\omega]$.
    By \eqref{eq: componentwise output}, $(H(y+\nu))_j=S_0(x_j)+(H\nu)_j\in[\omega/2,1-\omega/2]$, so that $\hat x_j$ is well defined and $|\hat x_j-x_j|\le\ell_\omega\delta_H\le c_\nu\delta$.
\end{proof}

\section{Numerical simulation} \label{sec: simulation}

We illustrate the procedure of Theorem~\ref{thm: full identification} on a two-dimensional Wilson--Cowan type network with piecewise linear activation \cite{Wilson1972,coombes2018networks},
\begin{equation}
    \begin{cases}
        \dot x=-\lambda x+CS(x)+\sigma_w\dot w+u, & \\
        y=WS(x)+\nu,                              &
    \end{cases}
\end{equation}
with $n=q=m=2$, $\lambda=1.2$, and $C$, $W$ unknown matrices whose entries are drawn independently and uniformly in $[-1,1]$.
The saturation is componentwise, $S(x)=(S_0(x_1),S_0(x_2))^\top$, with
\begin{equation}
    S_0(\xi)=\operatorname{clamp}\!\left(\frac{\xi+2}{4},\,0,\,1\right)=\begin{cases} 0 & \xi\le-2, \\[1mm] \dfrac{\xi+2}{4} & -2<\xi<2, \\[1mm] 1 & \xi\ge2. \end{cases}
\end{equation}
Here $w$ is a standard Brownian motion and $\nu$ a zero-mean Gaussian white noise, with $\sigma_w=0.5$ and $\sigma_\nu=0.3$ unless stated otherwise.
The system is integrated from $x(0)=(2,4)^\top$ by an Euler--Maruyama scheme with step $T_s=10^{-3}$, over $100$ runs in which $C$, $W$ and the noises are redrawn.
This is \eqref{eq: system-B-invertible} with $A=-\lambda\idty$ known and $\phi=CS(x)+\sigma_w\dot w$.
Since the noise $\sigma_w\dot w$ is not bounded, the simulation probes the method slightly outside the class covered by the theorems.

\begin{figure}[t]
    \centering
    \includegraphics[width=0.45\textwidth]{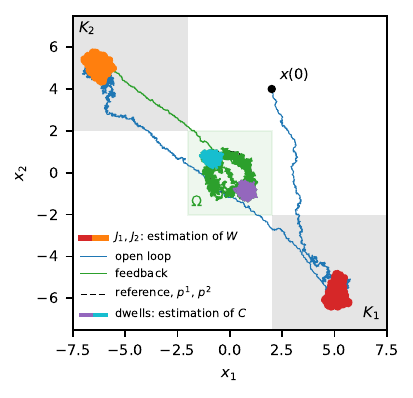}
    \hspace{0.1\textwidth}
    \includegraphics[width=0.34\textwidth]{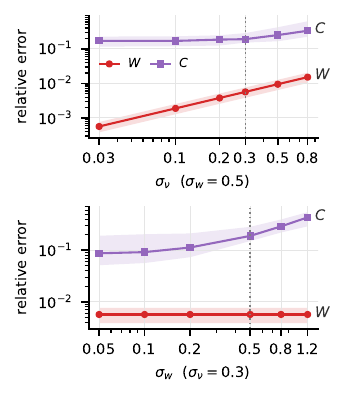}
    \caption{Left: trajectory of $x$ in the phase plane for one run.
            Right: relative errors $\|\hat W-W\|_F/\|W\|_F$ and $\|\hat C-C\|_F/\|C\|_F$, median and interquartile range over the runs, against $\sigma_\nu$ (top) and $\sigma_w$ (bottom).
            Dotted: the noise levels of the left panel.}
    \label{fig: Trajectory orthant}
\end{figure}

\paragraph{Identification of $W$}
The open-loop controls are taken along $b=(1,-1)^\top$, whose span meets the regions $K_1=[2,+\infty)\times(-\infty,-2]$ and $K_2=(-\infty,-2]\times[2,+\infty)$ of \eqref{eq: componentwise cone}, where $S$ takes the values $v_1=e_1$ and $v_2=e_2$.
With $\tau=10$, $t_1=\tau$ and $t_2=3\tau$, the control $u=(\mu_1\mathbf{1}_{[0,t_1+\tau]}-\mu_2\mathbf{1}_{[t_1+\tau,t_2+\tau]})b$, with $\mu_1=6$ and $\mu_2=7$, is the control of Lemma~\ref{lemma: sequential} for the design $\mathcal{E}=\{(v_1,[t_1,t_1+\tau]),(v_2,[t_2,t_2+\tau])\}$.
Its equilibria $\mu_1b/\lambda=(5,-5)$ and $-\mu_2b/\lambda\simeq(-5.8,5.8)$ lie inside $K_1$ and $K_2$ by more than the shift $1/\lambda\simeq0.83$ due to the coupling term.
Case \emph{ii.} of Theorem~\ref{thm: arbitrary time} would allow an arbitrarily short horizon, but long windows display the averaging of the output noise by \eqref{eq: estimator}.
Indeed, $\Gamma(\mathcal{E})=\tau\idty$, which implies that the root mean square error is $2\sigma_\nu(T_s/\tau)^{1/2}$ in Frobenius norm.
The bound of Remark~\ref{rem: noise identification}, which holds for any bounded noise, does not see this averaging.
The median error in Figure~\ref{fig: Trajectory orthant} (right) coincides perfectly with this prediction.

\paragraph{Tracking}
At $T=t_2+\tau=40$ the control switches to the output feedback \eqref{eq: two-stage control}, built with $\hat W=E(y;\mathcal{E})$ and fed with the noisy output, with $k=10$ and $\omega=0.1$.
On $[T,T+20]$ the reference is the circle $\gamma(t)=\bigl(\cos\frac{t-T}2,\sin\frac{t-T}2\bigr)^\top\in\Omega_{2\omega}=[-1.2,1.2]^2$.
On $[T+5,T+20]$ the state stayed in $\mathring\Omega$ in $99$ of the $100$ runs, and the tracking error $\|x(t)-\gamma(t)\|$, averaged over this interval, had a median of $0.33$.
It grows with $\|H\|_\infty=\|\hat W^{-1}\|_\infty$, from $0.21$ for $\|H\|_\infty<2$ to $0.60$ for $\|H\|_\infty\ge8$.
This holds although $\nu$ is unbounded and $k$ is below the gain $k^\star$ of that theorem.

\paragraph{Identification of $C$}
The reference then moves along a $C^1$ path, in $2$ time units, to $p^1=(0.8,-0.8)^\top$, dwells there for $100$ time units, and moves likewise to $p^2=-p^1$, where it dwells again.
Since $\lambda$ is known and $P=[S(p^1),S(p^2)]$ is invertible, this is item \emph{iii.} of Theorem~\ref{thm: full identification} with $\sigma=S$.
The estimate $\hat C:=[d_1,d_2]Z^{-1}$ of \eqref{eq: C regression} is evaluated with $\hat x=S_0^{-1}(Hy)$ in place of $x$ and $Hy$ in place of $S(x)$, where $H=\hat W^{-1}$ and $S_0^{-1}$ is extended\footnote{The extension is needed since the unbounded noise $\nu$ drives $Hy$ out of $[0,1]$.} affinely outside $[0,1]$, on windows $I_\ell$ starting $3$ time units into each dwell, with two changes against the noise.
The indicator of $I_\ell$ is replaced by a trapezoidal weight with ramps of $5$ time units, which averages the endpoint values of $\hat x$, and $\hat x$ in $d_\ell$ and $Hy$ in $\zeta_\ell$ are sampled at alternate steps, so that their errors are independent.
Without noise, $\hat C$ coincides with $C$ up to $10^{-5}$.
Figure~\ref{fig: Trajectory orthant} (right) shows that the error grows with both noises and is dominated by the state noise, which lies outside the class of Theorem~\ref{thm: full identification}.
This noise also drives the state out of $\mathring\Omega$ at some time after $T+5$ in $12$ runs.

\bibliographystyle{abbrv}
\bibliography{refs}

\end{document}